\documentclass[lettersize,journal]{IEEEtran}
\usepackage{amsmath,amsfonts}
\usepackage[linesnumbered,ruled]{algorithm2e}
\usepackage{algorithmic}
\usepackage{array}
\usepackage[caption=false,font=scriptsize,labelfont=sf,textfont=sf]{subfig}

\usepackage{textcomp}
\usepackage{stfloats}
\usepackage{url}
\usepackage{verbatim}
\usepackage{xcolor}
\usepackage{graphicx}
\usepackage{makecell}

\providecommand{\eg}{\emph{e.g.,} }
\def\BibTeX{{\rm B\kern-.05em{\sc i\kern-.025em b}\kern-.08em
    T\kern-.1667em\lower.7ex\hbox{E}\kern-.125emX}}
\usepackage{balance}
\begin{document}
\title{Learning-based Two-tiered Online Optimization of Region-wide Datacenter Resource Allocation}
\author{Chang-Lin Chen,
 Hanhan Zhou,
 Jiayu Chen,
 Mohammad Pedramfar,
 Tian Lan,
 Zheqing Zhu,
 Chi Zhou,
 Pol Mauri Ruiz,
 Neeraj Kumar,
 Hongbo Dong, and Vaneet Aggarwal
\thanks{C. Chen is with the Elmore Family School of Electrical and Computer Engineering, Purdue University, West Lafayette, IN 47907, USA. H. Zhou and T. Lan are with the Department of Electrical and Computer Engineering, The George Washington University, Washington, DC 20052, USA. J. Chen, M. Pedramfar, and V. Aggarwal are with the Edwardson School of Industrial Engineering, Purdue University, West Lafayette, IN 47907, USA. Z. Zhu, C. Zhou, P. M. Ruiz, N. Kumar, and H. Dong are with Meta Platforms, Inc., Menlo Park, CA 94025, USA. 

Contact: V. Aggarwal, vaneet@purdue.edu.

C. Chen, H. Zhou, M. Pedramfar, T. Lan, and V. Aggarwal acknowledge research award from Meta Platforms, Inc., and Office of Naval Research under grant N00014-23-1-2532.

This work is accepted to IEEE Transactions on Network and Service Management, Oct 2024. 

© 2024 IEEE. Personal use of this material is permitted. Permission from IEEE must be obtained for all other uses, in any current or future media, including reprinting/republishing this material for advertising or promotional purposes, creating new collective works, for resale or redistribution to servers or lists, or reuse of any copyrighted
component of this work in other works.
}}


\maketitle

\begin{abstract}
Online optimization of resource management for large-scale data centers and infrastructures to meet dynamic capacity reservation demands and various practical constraints (e.g., feasibility and robustness) is a very challenging problem. 
Mixed Integer Programming (MIP) approaches suffer from recognized limitations in such a dynamic environment, while learning-based approaches may face with prohibitively large state/action spaces. To this end, this paper presents a novel two-tiered online optimization to enable a learning-based Resource Allowance System (RAS). 
To solve optimal server-to-reservation assignment in RAS in an online fashion, the proposed solution leverages a reinforcement learning (RL) agent to make high-level decisions, e.g., how much resource to select from the Main Switch Boards (MSBs), and then a low-level Mixed Integer Linear Programming (MILP) solver to generate the local server-to-reservation mapping, conditioned on the RL decisions. We take into account fault tolerance, server movement minimization, and network affinity requirements and apply the proposed solution to large-scale RAS problems. To provide interpretability, we further train a decision tree model to explain the learned policies and to prune unreasonable corner cases at the low-level MILP solver, resulting in further performance improvement. 
Extensive evaluations show that our two-tiered solution outperforms baselines such as pure MIP solver by over $15\%$ while delivering $100\times$ speedup in computation.
\end{abstract}


\maketitle

\section{Introduction}


An important issue in managing large-scale data centers and infrastructures involves the efficient placement and real-time management of containers or virtual machines on servers~\cite{Google_Verma2015,Hindman2011,tang2020twine}. 
With the continuous expansion and increasing popularity of cloud computing, the demand for more advanced workload allocation techniques within shared datacenter infrastructures is escalating. This demand can reach several thousand for a specific region. Within a hierarchical architecture comprising datacenters, main switch boards (MSB), and racks, the number of servers of different types in a region can reach millions \cite{newell_ras_2021}. More importantly, the cluster manager must ensure a capacity guarantee despite infrastructure lifecycle events such as datacenter maintenance and hardware failures at various levels, including racks, MSBs, and datacenters \cite{Verbitski2017}. It is challenging to achieve optimal container-to-server assignments within such an extensive system to uphold capacity guarantees.



{\color{black}The implementation of capacity reservations is a widely adopted approach in datacenter management to ensure guaranteed resource availability for specific business units. These reservations account for workload constraints, varied hardware configurations, and service level objectives (SLOs) \cite{aws,google_web}. Despite their widespread use, there has been limited research on how reservations handle large-scale server failures while maintaining guaranteed capacity.
To address this challenge, the Resource Allowance System (RAS) was introduced in \cite{newell_ras_2021}. 
The system organizes servers into reservations, grouping them into logical clusters designated for specific business units, with failure handling accounted for at various hardware scopes. Containers are then placed within these reservations, ensuring that each container's workload requirements are met.
A key advantage of RAS is that server grouping is not a prerequisite for container placement, meaning containers can be efficiently placed without having to complete the server clustering first. This two-level design enhances management flexibility and reduces the number of candidate servers, streamlining the placement process and improving overall efficiency in resource allocation.}
However, it's worth noting that MIP-based approaches \cite{newell_ras_2021} primarily focus on an offline optimization of server-to-reservation mapping by considering system snapshots. It can be suboptimal for practical cluster systems that require maximizing long-term performance objectives in an online fashion, subject to various types of dynamism. 
While existing works have considered using deep reinforcement learning (DRL) for optimizing the long-term performance of cluster management, they consider direct container placement without explicitly accounting for instances of failure \cite{mao2019learning}. Moreover, directly applying DRL algorithms to a problem with a large action space is impractical.

This paper considers the problem of minimizing server movement, the largest failure domain, and resource spread imbalance at the rack and MSB levels under the constraints of capacity guarantee and network affinity requirements, which determine the extent of server allocation from the same datacenter. 
Motivated by the goal of achieving optimal results over the long run under dynamism and overcoming the limitation of RL approach for large-scale systems, we propose a two-level design that combines a DRL-based algorithm for online decision-making and a low-level MILP for reducing the action space of the DRL agent. 
At the first level, a DRL agent is trained to make sequential decisions for each reservation, which are then passed to a novel action converter. Specifically, the DRL agent's action affects the distribution of the servers gotten from the MSBs and the additional resources given to ensure capacity guarantee and the action converter comprising a softmax converter and an exponential converter transformed the agent's action into specific numbers of servers to reserve from the available MSBs.
In the second level, we introduce a low-level MILP model that generates the server-to-reservation mapping. This mapping is designed with a dual focus: firstly, to minimize the need for server relocation, which involves reassigning servers to different reservations, and secondly, to account for the risk of capacity unavailability due to potential failures at the rack level.
To ensure the reliability of our algorithm, we employ a decision tree model to learn from our algorithm's demonstrations. Specifically, we train the decision tree model to discern how the DRL agent decides the number of servers to allocate to a reservation based on the state-action pairs gathered through the execution of our trained RL model and the low-level MILP solver. This trained decision tree provides valuable insights into the decision-making process of the DRL agent. Furthermore, the decision tree model allows us to identify and address corner cases, thereby enhancing the performance of our proposed algorithm.

In the experiment section, we compare our algorithm with several baselines, including a MIP solver \cite{newell_ras_2021}, two complete RL approaches in which each element of its action decides whether a server is assigned to a reservation, and heuristic approaches. 
We further consider a few new two-tiered baselines that adopt low-level MILP as their second level while differing in their first level, determining the allocation of resources from the MSBs.
We evaluate the performance of our proposed algorithm under a simulation of a region of data centers which follows practical system statistics.
The results indicate that our proposed algorithms consistently outperform the competition, achieving the lowest overall cost with no constraint violations.
Specifically, in $30$ experiments across various percentiles, our algorithms demonstrate a remarkable $10\%$ to $15\%$ improvement in overall cost compared to the baselines. Moreover, they excel in providing the necessary additional resources for failure scenarios.
Furthermore, our proposed algorithm significantly surpasses the speed of the MIP solver by over $100$ times. This combination of superior performance and efficiency makes our approach a compelling choice for dynamic resource allocation in cloud computing systems.

\subsection{Summary of Contributions} \label{subsec: contributions}
The main contributions are as follows:

\noindent {\bf 1. } We introduce a dynamic cluster management system that takes into account the impact of infrastructure lifecycle events and the presence of heterogeneous hardware.

\noindent {\bf 2. } We formulate the Dynamic Resource Allowance System Optimization (DRASO) problem, which is subsequently optimized using a DRL-based algorithm.

\noindent {\bf 3. } We propose a two-tier DRL-based algorithm. At the MSB level, a collaboration between a DRL agent and our proposed action converter facilitates resource allocation for each reservation. Subsequently, a low-level MILP solver is employed to determine the server-to-reservation mapping based on outcomes from the DRL agent. We then enhance the reliability of our proposed algorithm by identifying corner cases through a decision tree model and implementing corrections within the low-level MILP solver.

\noindent {\bf 4. } Simulation results underscore the effectiveness of our approach. It showcases a 15\% enhancement in performance compared to the MIP solver in minimizing overall costs. Moreover, our approach executes 30 rounds of decision-making in just 26 seconds, whereas the MIP solver requires nearly an hour to accomplish the same task.

The rest of this paper is organized as follows. Section \ref{sec:system_model} presents the system model. Problem formulation and the algorithm for optimizing the dynamic cluster management system are presented in Section \ref{sec: problem_formulation} and Section \ref{sec: proposed_framework_design}. We provide numerical results in Section \ref{sec: simulation}. Finally, Section \ref{sec: conclusion} concludes this study.
\section{System Model}\label{sec:system_model}
This paper focuses on the dynamic resource allowance system (RAS), originally introduced in \cite{newell_ras_2021}. RAS abstracts the capacity of heterogeneous hardware and the demands of a workload through relative resource units (RRUs). These RRUs consider the varying performance of workloads on different hardware components such as CPUs, memory, flash storage, and GPUs. Additionally, RAS introduces the concept of a "reservation," which represents a logical cluster of servers providing capacity to the workload of a specific business unit.
RAS optimizes the mapping between servers and reservations based on various factors to ensure ample capacity available for the different business units. These factors include capacity requirements, potential failures at the MSB or rack level, datacenter maintenance events, the diverse resources of heterogeneous hardware, datacenter topology, and the specific requirements of workloads.
Nevertheless, the complexity of datacenter structures, the vast scale of capacity needed, and the diverse characteristics of workloads pose significant challenges when it comes to efficiently providing guaranteed capacity in a given region. This paper seeks to achieve optimal outcomes in a dynamic RAS environment over the long term.

\begin{figure}[t]
    \centering
    \includegraphics[width=0.485\textwidth]{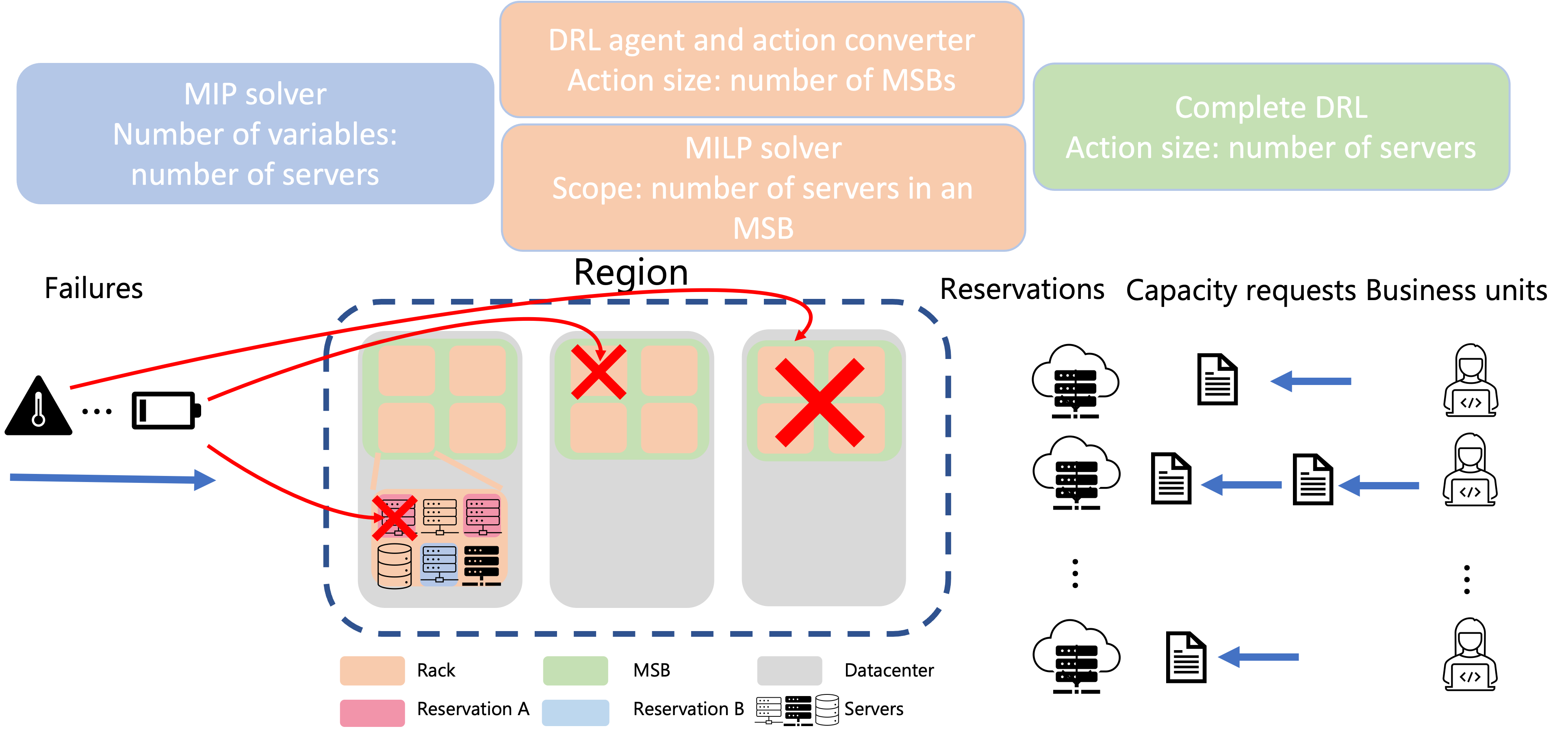}
    \caption{Depiction of dynamic Resource Allowance System considered in this paper, as well as a brief comparison of our two-tier approach, MIP solver, and the complete RL approach. The capacity demand of a reservation is an aggregation of the capacity requests over time. }
    \label{fig:RAS}
\end{figure}

In Fig. \ref{fig:RAS}, we show a region's layout overview. The datacenters in a region are connected by high-bandwidth network links. The main switch boards (MSBs) are isolated power and network domains that can fail independently. Failures occurring at this level are typically attributed to common physical devices such as power breakers and cooling fans. Additionally, failures can arise at the server or rack level, often stemming from hardware issues that necessitate repair. We denote $D$ as the set of datacenters, $F$ as the set of MSBs, $K$ as the set of racks, $B$ as the set of servers, and $L$ as a set of reservations. Considering hardware heterogeneity, the servers can be categorized into $E$ types. In \cite{newell_ras_2021}, users in a business unit send capacity requests to RAS, telling RAS how many additional resources they need. Then, RAS collects each reservation's capacity requests and periodically optimizes the mapping of servers and reservations.

In the system, time is divided into multiple time slots with equal duration $\Delta t$. Let $\mathbb{T} = \{1,2,\cdots,t,\cdots,\mathcal{T}\}$ be the set of time slots. 
Let $\mathbf{T}_l=\{\tau_{l,1}, \tau_{l,2}, \cdots\}$ be the set of capacity requests arriving for reservation $l$. We denote the arrival time of the capacity request $\tau_{l,i}$ as $t_{l,i}$, its capacity demand as $\mathbf{c}_{l,i}=(c_{l,i,1}, \cdots, c_{l,i,E})$ where $c_{l,i,e}$ is the demand of $\tau_{l,i}$ for server type $e$, and its expiration time as $t^{'}_{l,i}$. The reservations' demand varies with time as the capacity requests come and leave. We denote $\mathbf{C}_l(t)$ as the total capacity demand of the reservation $l$ at time $t$, and it is given as
\begin{equation} \label{eq: reservation_cap_calc}
    \mathbf{C}_l(t) = \sum_{i\in\Theta_t}\mathbf{c}_{l,i}-\sum_{j\in\Theta^{'}_t}\mathbf{c}_{l,j},
\end{equation}
where $\Theta_t=\{i| t_{l,i}\leq t\}$ and $\Theta^{'}_t=\{i| t^{'}_{l,i}\leq t\}$ are the indices of the capacity requests arriving and expiring right before time $t$, respectively. Note that $\mathbf{C}_l(t)=(C_{l,1}(t), \cdots, C_{l,E}(t))$, where $C_{l,e}(t)$ is the capacity demand for the type $e$ servers. According to the operation of RAS, the demand and supply of servers of one type are independent of another type. 

We use a one-hot vector $\mathbf{h}_b=(h_{b,1}, \cdots, h_{b,E})$ to indicate which type server $b$ is. 
Let $U_{b}$ be the RRU of server $b$ and $x_{b,l}(t)$ be a binary variable equal to $1$ if server $b$ is assigned to reservation $l$. 
\paragraph{Server Assignment Constraint}
Assuming a server can only be assigned to one reservation at a time, servers at time slot $t$ should satisfy the following constraint,
\begin{equation} \label{eq: constr_server_assignment}
    g_{1,b}(t) = 1-\sum_{l\in L}x_{b,l}(t)\geq 0, \forall b \in B.
\end{equation}
The objectives of RAS are to minimize server movement and optimize server spread across fault domains of various scopes. 
\paragraph{Server Movement Cost}
When RAS assigns a server to another reservation, the overhead of moving containers running on the server burdens the system. 
The movement cost of server $b$ at time $t$ is given as
\begin{equation}\label{eq: movement_cost_type}
    o_{1,b}(t) = \sum_{l=1}^{|L|}M_b\times\max(0, x_{b,l}(t-1)-x_{b,l}(t)),
\end{equation}
where $M_b$ is the movement cost of server $b$. 
\paragraph{Outside Rack and MSB Spread Goals}
For fault tolerance purposes, RAS typically assigns more capacity than the reservations ask for, and aims to minimize 
the amount of capacity that exceeds the desired proportion of the capacity demand for each reservation and server type,
\begin{equation}\label{eq: rack_spread}
    \begin{aligned}
    o_{2,k,l,e}(t) &= \sum_{G\in\Psi^K}\max(0, \sum_{b\in G}(h_{b,e}\times U_{b}\times x_{b,l}(t))-\\
    &\alpha^K\times {C}_{l,e}(t)),
    \end{aligned}
\end{equation}
and
\begin{equation}\label{eq: msb_spread}
    \begin{aligned}
        o_{3,f,l,e}(t) &= \sum_{G\in\Psi^F}\max(0,\sum_{b\in G}(h_{b,e} {U}_{b} x_{b,l}(t))-\\
        &\alpha^F\times{C}_{l,e}(t)),
    \end{aligned}
\end{equation}
where $\Psi^K$ and $\Psi^F$ are partitions of servers based on racks and MSBs, and $\alpha^K$ and $\alpha^F$ are the spread parameters that decide the threshold for fulfilling the capacity demand within a physical scope.
Making $\alpha^K$ and $\alpha^F$ small minimizes the effect of failures at the racks and MSBs levels on a reservation while making them large gives RAS more flexibility to optimize other objectives. Specifically, when $\alpha^K$ and $\alpha^F$ are small, the optimal resource supply tends to be more evenly distributed across all racks and MSBs. Conversely, larger values of $\alpha^K$ and $\alpha^F$ allow for non-uniform resource allocation among scopes, thereby enabling consideration of more solutions to minimize the overall objective function.
\paragraph{Cost of Largest Failure Domain}
To minimize the impact of failures at the MSB level, RAS also minimizes the largest amount of resources reservation $l$ can get from the MSBs for server type $e$,
\begin{equation}\label{eq: largest_msb}
    o_{4,l,e}(t) = \max_{G\in\Psi^F}\left(\sum_{b\in G}h_{b,e}\times U_{b}\times x_{b,l}(t)\right),
\end{equation}
for all $l$ and $e$. 

\paragraph{Capacity Guarantee}
Moreover, to guarantee that the supply of servers of a type is enough to satisfy the capacity demand even if the MSB that provides the most capacity fails, the server-to-reservation mapping generated by RAS should satisfy the following constraint,
\begin{equation}\label{eq: capacity_redundancy}
    \begin{aligned}
        g_{2,l,e}(t)&=\sum_{b\in B}(h_{b,e}\times U_{b}\times x_{b,l}(t))-\\
        &\max_{G\in\Psi^F}\left(\sum_{b\in G}h_{b,e}\times U_{b}\times x_{b,l}(t)\right) - C_{l,e}(t)\geq 0,
    \end{aligned}
\end{equation}
for all $l$ and $e$.
\paragraph{Network Affinity Requirements}
Considering that business units require different extents of network affinity, representing the extent to which a reservation's resource allocation is concentrated across data centers, RAS uses the following constraint,
\begin{equation}\label{eq: network_affinity}
    g_{3,d,l,e}(t)=\theta-\left|\frac{\sum_{b\in \Psi^D_d}(h_{b,e}\times U_{b}\times x_{b,l}(t))}{{C}_{l,e}(t)}-{A}_{d,l,e}\right|\geq 0,
\end{equation}
to enforce a reservation's preference for physical datacenters. In \eqref{eq: network_affinity}, the term $\left|\frac{\sum_{b\in \Psi^D_d}(h_{b,e}\times U_{b}\times x_{b,l}(t))}{{C}_{l,e}(t)}-{\bf A}_{d,l,e}\right|$ represents the proportion of resources from datacenter $d$ relative to the reservation's demand, where $A_{d,l,e}$ signifies the desired ratio. $\theta$ determines the acceptable deviation from the desired ratio.
The RAS updates the server-to-reservation assignment at the end of each time slot, considering the above objectives and constraints. We summarize the notions in Table \ref{table: op_param_summary}.  The following section shows the optimization problem RAS considered and how it is optimized.
\section{Problem Formulation} \label{sec: problem_formulation}
This section presents the problem considered in this work.
RAS formulates the utility function considering the churn rate and allocation spread across MSBs and racks {\color{black}of a snapshot of the system} as,
\begin{equation} \label{eq: total_utility}
    \begin{aligned}
        \mathcal{U}(t)=&\sum_{b\in B}o_{1,b}(t)+\sum_{e=1}^E\sum_{l\in L}\Big(\beta\times\big(o_{2,l,e}(t)+o_{3,l,e}(t)\big)\\
        &+\kappa\times o_{4,l,e}(t)\Big),
    \end{aligned}
\end{equation}
where $\beta$ and $\kappa$ are the cost of capacity out of spread goals and the cost of correlated failure buffer capacity. {\color{black}$M_b$ in (3), along with $\beta$ and $\kappa$, are defined by cloud providers providing trade-offs among server movement, resource distribution across different scopes, and impact of the largest failure domain [4]. These parameters can be adjusted by the system designer to achieve overall SLOs.}

In this work, we aim to obtain long-term optimal {\color{black}datacenter management} under dynamism, considering the constraints of capacity guarantee and the network affinity requirements for each reservation.
We define our Dynamic Resource Allowance System Optimization (DRASO) problem and present it as:
\begin{align} \label{eq: optimization_problem}
    \begin{split}
    \text{DRASO}:&\\
    \min_{\substack{x_{b,l}(t),\forall l\in L,\\ b \in B, \mathcal{T}\geq t\geq 0} } &\sum_{t=0}^{\mathcal{T}}\gamma^{\mathcal{T}-t}\mathcal{U}(t)
    \end{split}\\
    s.t.\quad
    \begin{split}
        &g_{1,b,e}(t) = 1-\sum_{l\in L}x_{b,l}(t)\geq 0,\\
        &\quad\quad\quad\quad\quad\forall b \in B, \text{ and }\mathcal{T}\geq t\geq 0,
    \end{split}\\
    &g_{2,l,e}(t) \geq 0, \forall l \in L, \text{ and }\mathcal{T}\geq t\geq 0,\\
    &g_{3,d,l,e}(t)\geq 0, \;\; \forall d \in D, l\in L, \text{ and }\mathcal{T}\geq t\geq 0,
\end{align}\label{prob: original_problem}
where $\gamma \in [0,1)$ is a discounting factor.
The above problem is a mixed integer programming (MIP) problem. Note that the number of variables in DRASO can go beyond millions for a region given a time slot $t$. Moreover, the capacity requests which are aggregated into $C_{l,e}(t)$ and the assignment of servers can stay in the system and impact the decision-making for the next few rounds of system optimization. RAS uses the MIP solver to optimize the server-to-reservation mapping at a region level at a given $t$ and, thus, will not give an optimal online solution. To achieve long-term optimal results, we consider adopting Reinforcement Learning (RL), widely adopted to solve sequential decision-making problems in a dynamic environment. Moreover, considering the number of variables can be very large, we combine an action converter and a low-level MILP with DRL to improve efficiency. The following section will detail each component in our proposed algorithm.

\section{Proposed Framework Design} \label{sec: proposed_framework_design}
In this section, we present our proposed framework and hybrid DRL-based approach to solving DRASO. Our framework comprises an agent, an environment, an action converter, and a low-level MILP. In RL, an agent is trained to get long-term optimal results by sequentially evaluating its decisions sent to the environment representing our target system. The action converter and the low-level MILP are components of our approach to efficiently solving the problem.
We begin with a comprehensive framework overview in Section \ref{subsec: framework} and subsequently detail each component in Sections \ref{subsec: ppo}, \ref{subsec: action_converter}, and \ref{subsec: heuristic}. 
\subsection{Framework} \label{subsec: framework}
\begin{figure}[t]
    \centering
    \includegraphics[width=0.485\textwidth]{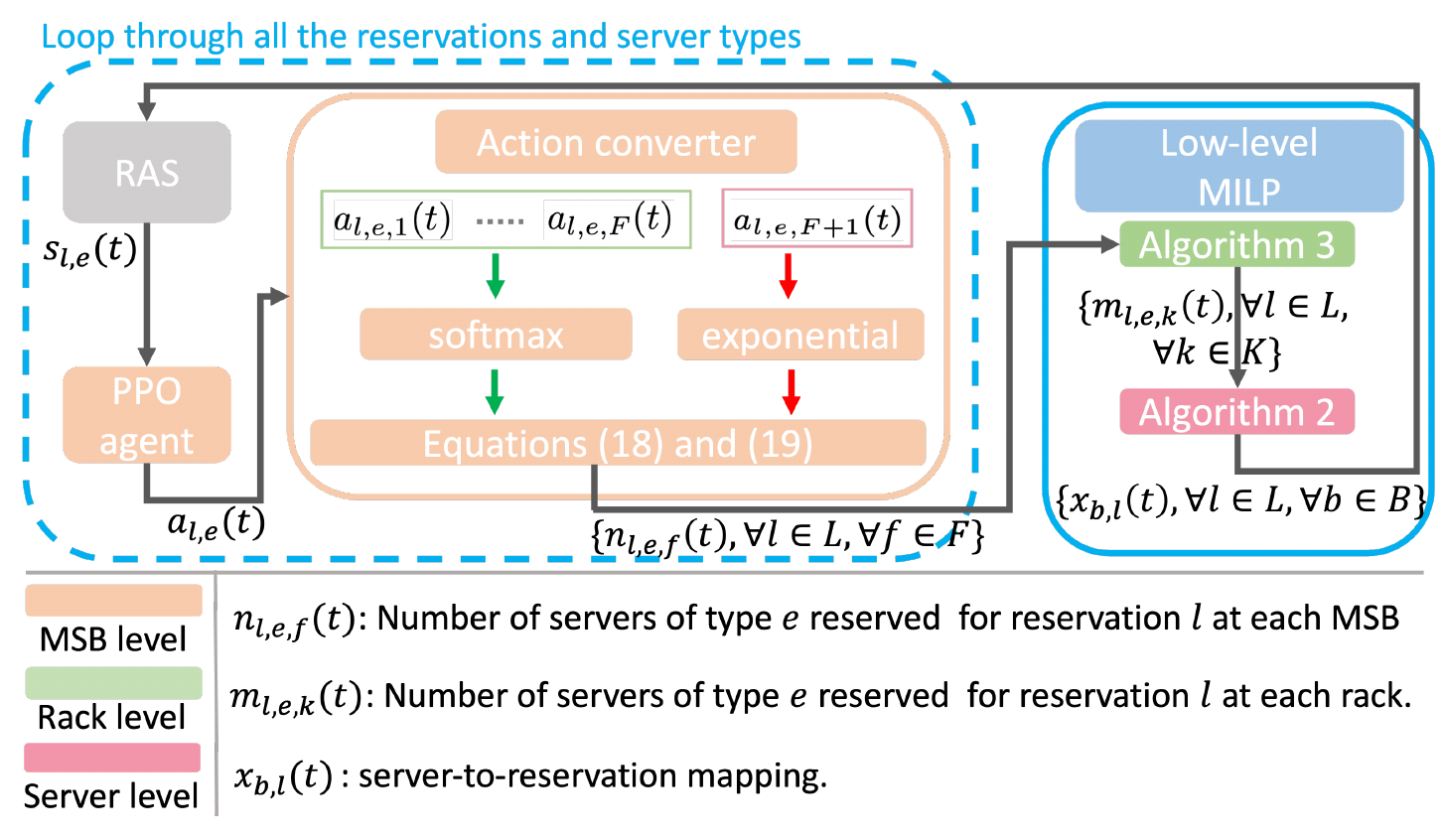}
    \caption{A description of our proposed framework, where the PPO agent makes decisions for every reservation and then transforms into a number of servers to take from the MSBs by the action converter. Finally, the number of servers to take from the MSBs for all reservations is converted into server-to-reservation mapping by the low-level MILP.}
    \label{fig: framework}
\end{figure}
We start with dividing the system to decide where to apply the DRL algorithm.
We find that the utility function $\mathcal{U}(t)$ in DRASO can be decoupled based on the server types and the utility function for server type $e$ can be expressed as
\begin{equation} \label{eq: utility_server_type}
    \begin{aligned}
        \mathcal{U}_e(t)=&\sum_{b\in B_e}o_{1,b}(t)+\sum_{l\in L}\Big(\beta\times\big(o_{2,l,e}(t)+o_{3,l,e}(t)\big)\\
        &+\kappa\times o_{4,l,e}(t)\Big),
    \end{aligned}
\end{equation}
where $B_e$ is the set of type $e$ servers. Moreover, as mentioned in Section \ref{sec:system_model}, the demand of a reservation is an aggregation of multiple capacity requests that ask for multiple types of servers and can arrive and expire. As shown in Eq. \eqref{eq: reservation_cap_calc}, each tuple of the capacity demand of a reservation represents the demand for a server type, and satisfying each tuple is independent according to the RAS operation. Therefore, the utility function and the constraints can be fully decoupled according to the server types, and the optimization can be done separately for each server type.

The proposed approach for solving the optimization problem DRASO in \eqref{eq: optimization_problem} consists of two levels.
In the first level, the agent is responsible for making decisions, which are then passed to the action converter to transform them into the actual number of servers to be allocated from the MSBs.
In the second level, a low-level MILP is employed to determine the server-to-reservation mapping, taking into account the results obtained from the first level for all reservations and the current state of the environment.
As discussed in the conclusion of Section \ref{sec: problem_formulation}, the rationale behind the two-layer design stems from the large number of variables involved.

Fig. \ref{fig: framework} shows the basic blocks of our proposed framework and communications between them. The environment maintains the states of the servers, reservations, and capacity requests, e.g., the capacity of the reservations, previous server-to-reservation mapping, etc. These states are updated at every time slot based on the arrival and expiration of the capacity requests and the mapping decisions. 
\begin{algorithm}[t]
  \DontPrintSemicolon
  \SetAlgoLined
  \KwIn{\{$D$, $F$, $K$, $B$, $L$, $\mathcal{T}$\} }
  \KwOut{$\{x_{b,l}(t), \forall b\in B, \forall l\in L\}$}
  $t \gets 1$\;
  \Repeat{$t==\mathcal{T}$}{
    \For{$e=1$ \KwTo $|E|$}{
    \For{$l=1$ \KwTo $|L|$}{
    Send $\mathbf{s}_{l,e}(t)$ to the trained PPO agent and get $\mathbf{a}_{l,e}(t)$\;
    Convert $\mathbf{a}_{l,e}(t)$ to numbers of servers to take by reservation $l$, $n_{l,e,f}(t), \forall f \in F$, by \eqref{eq: num_servers_get_msb}\;
    Get numbers of servers to take from the rack $k$, $m_{l,e,k}(t), \forall k\in K$, by Algorithm \ref{alg: even_servers_from_racks}\;
    }
    Get server-to-reservation mapping, $\{x_{b,l}(t), \forall b\in B_e, \forall l\in L\}$ by Algorithm \ref{alg: actual_mapping} and send it to the environment\;
    }
    $t \gets t+1$\;
    }
  \caption{Proposed algorithm\label{alg: hirarchical_drl_alg}}
\end{algorithm}

\subsection{PPO Agent} \label{subsec: ppo}
In this section, we reframe RAS as a Markov Decision Process (MDP) that can be effectively addressed using readily available RL algorithms. For this study, we've selected PPO \cite{DBLP:journals/corr/SchulmanWDRK17} as the algorithm because it requires less hyperparameter tuning to get satisfactory results, as well as it can generate continuous actions, and it is simple, efficient, and stable. An MDP comprises key elements such as the state space, action space, and reward function. We present their definitions in the context of RAS as follows:
\begin{enumerate}
    \item \emph{State}: The state variables are defined to reflect the environment status and thus affect the reward feedback of different actions. We denote $\mathbf{s}_{l,e}(t)$ as the state variable sent to the agent to decide for the reservation $l$ and the server type $e$. The state elements include the index of the reservation, the usage of the MSBs at the current time slot, the capacity demand of the reservation $l$, the number of servers in each MSB previously assigned to the reservation $l$, the amount of demand that will expire in the future $h$ slots, and the amount of demand that will arrive in the future $h$ time slots. The demand arrivals and expiration in the future can be predicted by an LSTM model trained by historical data. In addition, the states include the mean of the other reservations' capacity demand, the number of previously assigned servers from the MSBs, to-be expired demand in the future $h$ time slots, and demand arrivals in the future $h$ time steps. The state elements are combined in one vector $\mathbf{s}_l(t)$. 
    \item \emph{Action}: 
    We denote the the agent's output for reservation $l$ and server type $e$ as $\mathbf{a}_{l,e}(t)=(a_{l,e,1}(t)$, $\cdots$, $a_{l,e,F}(t)$, $ a_{l,e,F+1}(t))$ where $\mathbf{a}_{l,e}(t)\in \mathbb{R}^{F+1}$. $\mathbf{a}_{l,e}(t)$ will be sent to the action converter. At the action converter, $a_{l,e,1}(t)$, $\cdots$, $a_{l,e,F}(t)$ will first be converted to the fraction of capacity demand to take from the MSBs, and $a_{l,e,F+1}(t)$ will be converted to an over-provision factor.
    Then, the action converter will further derive the numbers of servers of type $e$ to take from the MSBs. 
    $a_{l,e,1}(t)$, $\cdots$, $a_{l,e,F}(t)$ affect whether the servers taken are evenly from the MSBs. $a_{l,e,F+1}(t)$ determines the additional servers for the failure buffer.
    The conversion is detailed in Section \ref{subsec: action_converter}.
    \item \emph{Reward}: After converting the agent's action into a server-to-reservation mapping, the reward for the reservation's action is determined by a weighted combination of the equations in the objective of the optimization problem in \eqref{eq: optimization_problem} plus penalties for violation of the constraints. The reward for reservation $l$ at time slot $t$ is given by
    \begin{equation}
        \begin{aligned}
            r_{l,e}(t) &= r(s_{l,e}(t), a_{l,e}(t)) = w_1\times\sum_{b\in B_e}o_{1,b}(t)+\\
            &w_2\times\sum_{k\in K}o_{2,k,l,e}(t)+w_3\times\sum_{f\in F}o_{3,f,l,e}(t)+\\
            &w_4\times o_{4,l,e}(t)+p_{2,l,e}\times \mathbf{1}\{g_{2,l,e}(t)<0\}+\\
            &\sum_{d\in D}p_{3,d,l,e}\times \mathbf{1}\{g_{3,d,l,e}(t)<0\}
        \end{aligned}
    \end{equation}\label{eq: reward_reservation}
    where $w_1$, $w_2$, $w_3$, and $w_4$ are the weights for the four equations related to reservation $l$ in the objective, $p_2$ and $p_{3,d}$ are penalties for constraint violation, and $\mathbf{1}\{\cdot\}$ is an indicator function equal to $1$ if the statement in the bracket holds. The penalty for violating constraints $g_{1,b}(t)$ is not included in the reward function because the proposed low-level MILP (introduced later) takes care of it. 
    To enhance training efficiency, curriculum learning \cite{curriculum_ronan} strategy is adopted. The agent is initially trained using only the first term of the reward function. Once the agent's performance converges, the next term is added, and the agent is retrained. This process is repeated incrementally until the agent is trained with the complete reward function.
    By gradually introducing the reward components, the training process becomes more efficient, allowing the agent to focus on learning each term sequentially. This approach helps in achieving convergence and improving the overall training performance.
\end{enumerate}

The PPO agent takes actions sequentially for each reservation and server type according to the environment's state and learns from the records of actions, states, and rewards. The order of the decision-makings for the reservations matters in our design. The later reservations might not get enough resource to satisfy its demand. To mitigate the effect, we introduce the reservation index and the mean of the other reservations' states into the state variables, so that the agent is aware of which reservation it is making the decision for and can take decisions made for other reservations into account \cite{mondal2022can}.
On the contrary, the optimization of server assignment for the server types is independent of each other, so we propose the following two training methods.
\begin{itemize}
    \item Single agent: We train a single agent for all the server types. To achieve this, we add the server type to the states. Following the concept of contextual MDP, we then train the agent to solve the optimization problem for a server type and move on to the next one after finishing the current training \cite{hallak_contextual_2015}.
    \item Parallel agents: We train $|E|$ agents, and each is for a server type.
\end{itemize}
The Parallel agents enable faster training by parallelizing the training workload to multiple machines.
We summarize the training of the PPO agent in Algorithm \ref{alg: ppo_train} in Appendix \ref{apdx: algorithms}.

\subsection{Action Converter} \label{subsec: action_converter}

In our system design, the PPO agent's output undergoes a sequential procedure involving our proposed action converter and low-level MILP to generate the requisite server-to-reservation mapping.

The purpose of the action converter is to determine the number of servers a reservation should get from every MSB given the actions from the PPO agent. 
The conversion involves converting the action to fractions of capacity to take from the MSBs and an over-provision factor and then getting the number of servers to take from the MSBs.
We pass $(a_{l,e,1}(t)$, $\cdots$, $a_{l,e,F}(t))$, to the softmax function to get the $(a^{'}_{l,e,1}(t)$, $\cdots$, $a^{'}_{l,e,F}(t))$ as follows
\begin{equation} \label{eq: fraction_cap}
    a^{'}_{l,e,f}(t) = \frac{e^{\zeta a_{l,e,f}(t)}}{\sum_{f\in F}e^{\zeta a_{l,e,f}(t)}}, \forall f \in F,
\end{equation}
where $\zeta$ is the temperature parameter that affects the uniformity of $(a^{'}_{l,e,1}(t)$, $\cdots$, $a^{'}_{l,e,F}(t))$.
We get the over-provision factor, $z_{l,e}(t)$, by $a_{l,e,F+1}$ and the following equation,
\begin{equation}\label{eq: over_provision}
    z_{l,e}(t) = \omega^{a_{l,e,F+1}(t)},
\end{equation}
where $\omega$ is manually set.
The amount of type $e$ resource the reservation $l$ should get from the MSB $f$ is given as: 
\begin{equation} \label{eq: resource_get_msb}
    y_{l,e,f}(t) = a^{'}_{l,e,f}(t)\times(1+z_{l,e}(t))\times C_{l,e}(t).
\end{equation}
Then, the number of type $e$ servers that the reservation $l$ gets from the MSB $f$ should be:
\begin{equation} \label{eq: num_servers_get_msb}
    n_{l,e,f}(t) = y_{l,e,f}(t)*\frac{|B_e|}{\sum_{b\in B_e}U_{b,l}}.
\end{equation}
By \eqref{eq: fraction_cap}, \eqref{eq: over_provision}, \eqref{eq: resource_get_msb} and \eqref{eq: num_servers_get_msb}, the action converter gets the number of servers a reservation should get from the MSBs.

\begin{algorithm}[tbhp]
  \DontPrintSemicolon
  \SetAlgoLined
  \KwIn{$\{m_{l,e,k}(t), \forall l\in L, \forall k\in K\}$}
  \KwOut{$\{x_{b,l}(t), \forall l\in L, \forall b\in B_e\}$}
  Loop through the racks to get the difference between the total requests and the capacity of the rack, $u_{e,k}(t) \gets \sum_{r=1}^Rm_{l,e,k}(t)-|\Psi^K_k|$\;
  Loop through the racks and collect the requests to make the total requests less than the rack capacities\;
  Loop through the collected requests to first reassign them to the racks in the same MSB they were sent to\;
  Assign the requests that cannot be fulfilled in the previous step to any rack having vacancies\;
  Loop through the servers to get server-to-reservation mapping\;
  \caption{Description of how to get the server-to-reservation mapping for all the reservations}\label{alg: actual_mapping}
\end{algorithm}

\subsection{Integration with the Low-level MILP} \label{subsec: heuristic}
With the results from the action converter, our low-level MILP derives the server-to-reservation mapping. Since the resource spread across the MSBs and the maximum MSB failure are handled by the PPO agent, our low-level MILP is designed to minimize the movement cost of servers in \eqref{eq: movement_cost_type} and the resource spread across the racks in \eqref{eq: rack_spread}. The essential steps in our low-level MILP to get the server-to-reservation mapping are as follows:
\begin{enumerate}
    \item Given the number of servers the reservation $l$ should get from the MSB $f$, $n_{l,e,f}(t)$, the reservation $l$ gets servers evenly from the racks in the MSB $f$. We use Algorithm \ref{alg: even_servers_from_racks} to decide how many servers the reservation $l$ should get from all the racks.
    \item Based on the number of servers every reservation should take from the MSBs, we can get the server-to-reservation mapping. The detailed steps are shown in Algorithm \ref{alg: actual_mapping}, and the corresponding pseudocode can be found in Algorithm \ref{alg: actual_mapping_details} in Appendix \ref{apdx: algorithms}.
\end{enumerate}

The time complexity of Algorithm \ref{alg: hirarchical_drl_alg} is $\mathcal{O}(|E||L||K|+|B|)$ since the time complexities of Algorithm \ref{alg: actual_mapping} and Algorithm \ref{alg: even_servers_from_racks} are $\mathcal{O}(|B_e|)$ and $\mathcal{O}(|K|)$, respectively.

\section{Simulation} \label{sec: simulation}
We evaluate the performance of our proposed algorithm in this section. We begin with the system setup in Section \ref{subsec: system_setup}, clarify the baselines in Section \ref{subsec: baselines}, and show the performance comparison in Section \ref{subsec: main_results}.
\subsection{System Setup}\label{subsec: system_setup}

The counts of datacenters, MSBs, racks, reservations, servers, and server types are symbolized as $|D|$, $|F|$, $|K|$, $|L|$, $|B|$, and $|E|$, respectively. These symbols illustrate the regional datacenter topology depicted in Fig. \ref{fig:RAS}. $U_b$, representing the RRU of server $b$, is unified since most of the servers of the same server type are homogeneous. This datacenter setup is designed by domain experts among the authors, approximating a real-life scenario. For one experiment/episode, we run the system for $30$ steps. The parameters  related to the optimization problem are 
$|D|=3$, $|F|=15$, $|K|=75$, $|L|=20$, $|B|=1000$, $|E|=10$, $U_b=150$, $T=30$, $\alpha^F=1/15$, $\alpha^K=1/75$, $\kappa=1$, $\beta=1$, $M_b=5$, $A_{l,d}=1$, and $\theta=2$, where the definitions of these are provided in Table \ref{table: op_param_summary}.

The regional datacenter comprises $1000$ servers, differentiated into $10$ types based on count, mean request arrival rate, and capacity combination type. {\color{black}The arrival of the capacity request is assumed to follow a Poisson process with rate $\lambda_l$ at server $l$.} A comprehensive summary of this heterogenous simulation setting, designed by domain experts, is provided in Table \ref{table: 2}.

\begin{table}[htbp] 
\small
\caption{Summary of the Heterogeneous Hardware} \label{table: 2}
\centering
\resizebox{\columnwidth}{!}{%
 \begin{tabular}{| c || c | c | c | c | c | c | c | c | c | c |} 
 \hline
 Server Type & 0 & 1 & 2 & 3 & 4 & 5 & 6 & 7 & 8 & 9 \\
 \hline\hline
Count & 405 & 45 & 30 & 15 & 300 & 15 & 30 & 15 & 45 & 100 \\
 \hline
 \makecell[c]{Mean Request \\ Arrival Rate} & 2 & 0.6 & 0.6 & 0.2 & 1.8 & 0.2 & 0.6 & 0.2 & 0.6 & 1.2 \\
 \hline
 \makecell[c]{Capacity \\ Combination Type} & 5 & 2 & 1 & 0 & 4 & 1 & 2 & 1 & 2 & 3 \\
 \hline
 \end{tabular}
 }
\end{table}

There are in total $6$ types of capacity combinations. Each type corresponds to a unique amalgamation of demands varying in size or duration, summarized in Table \ref{table: 3}. The hardwares are evenly distributed in their higher scopes \eg the racks have the same number of servers, and the MSBs have the same number of racks. We build a simulator in Python based on the described settings.

\begin{table}[htbp] 
\small
\caption{Summary of Different Capacity Requests} \label{table: 3}
\centering
\resizebox{\columnwidth}{!}{%
 \begin{tabular}{| c || c | c | c | c | c | c |} 
 \hline
 \makecell[c]{Capacity \\ Combination Type} & 0 & 1 & 2 & 3 & 4 & 5 \\
 \hline\hline
Demands & (150) & (300, 150) & (300, 150) & (300, 150) & (450, 300) & (900, 450, 150)\\
 \hline
Percentage & (1.00) & (0.16, 0.84) & (0.25, 0.75) & (0.33, 0.67) & (0.33, 0.67) & (0.16, 0.33, 0.51) \\
 \hline
Duration & (15) & (10, 15) & (10, 15) & (10, 15) & (10, 15) & (10, 15, 20)\\
 \hline
 \end{tabular}
 }
\end{table}

\subsection{Baseline Algorithms} \label{subsec: baselines}

This section describes the baseline algorithms. We include the MIP solver, which RAS currently adopts, two complete RL agents, and some two-level heuristic methods. The descriptions of the baseline algorithms are given as follows. 
\begin{itemize}
    \item MIP solver: It is implemented by the Python-MIP package \cite{python_mip}. The MIP solver is configured to spend at most $10$ seconds to solve a problem.
    \item Complete PPO: The action space of the agent is equal to the set of variables in \eqref{eq: utility_server_type}, where each output of the agent is then converted to $0$ or $1$ to represent $x_{b,l}(t)$. We use the PPO algorithm to train the agent.
    \item Complete TD3: We use the TD3 algorithm \cite{td3_fujimoto} to train the agent which decides the value of the set of variables in \eqref{eq: utility_server_type}.
    \item Random: It uniformly samples $|F|$ weights and an over-provision factor from the range $[-1, 1]$. Then, the $|F|+1$ values are sent to the action converter and our low-level MILP to get the server-to-reservation mapping.
    \item Uniform: Let the fractions of capacity demand to take from the MSBs be $\frac{1}{|F|-1}$. The choice of the fraction will make \eqref{eq: capacity_redundancy} satisfied. The fractions are sent to the action converter and our low-level MILP to obtain the server-to-reservation mapping.
    \item Proportional: It first generates $F$ scalars, $\delta_1$, $\cdots$, $\delta_F$, proportional to the available resource of the MSBs. Assume that $(\delta_1$, $\cdots$, $\delta_F)$ are in the descending order and they sum to one. Then, multiply all the scalars by $\frac{1}{1-\delta_1}$ to satisfy \eqref{eq: capacity_redundancy}. Finally, the scalars are sent to the action converter and our low-level MILP to get the server-to-reservation mapping.
\end{itemize}

\subsection{Main Results} \label{subsec: main_results}
\begin{figure*}[htbp]
    \centering
    \subfloat[Objective Value Empirical CDF]{\label{fig:obj_cdf} \includegraphics[scale=0.37]{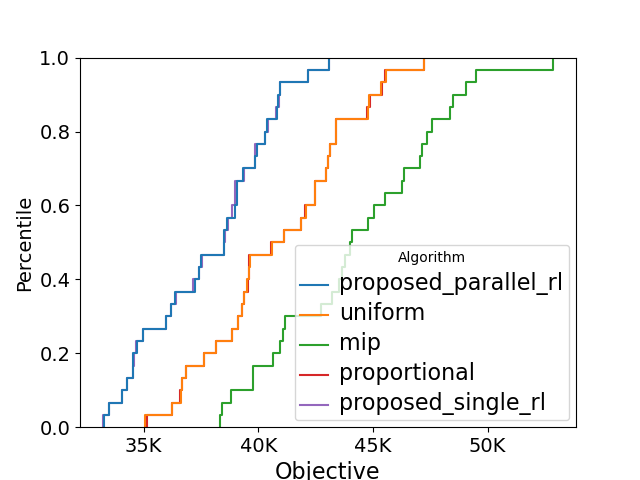}}
    \subfloat[Objective Value Empirical CDF]{\label{fig:obj_cdf_vanilla} \includegraphics[scale=0.37]{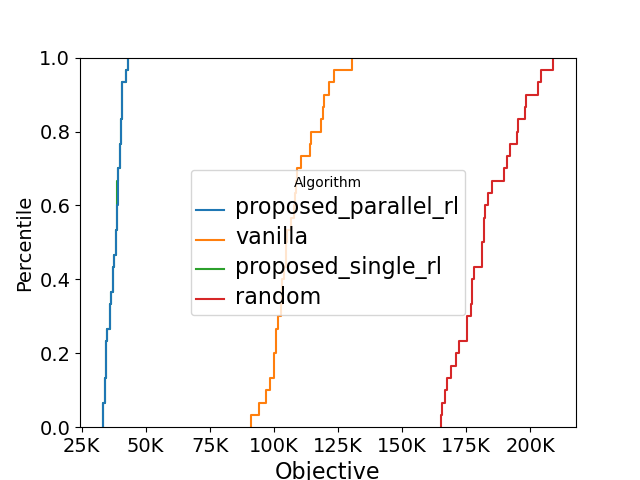}}
    \subfloat[Objective Value Variation]{\label{fig:obj_step} \includegraphics[scale=0.37]{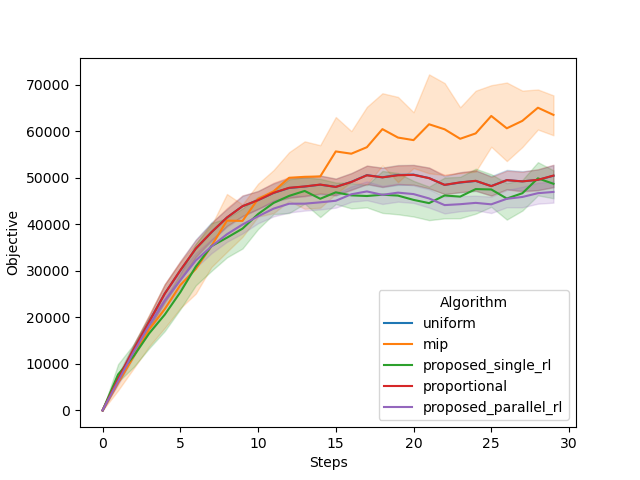}}
    \caption{Performance comparison with regards to the objective value ($\mathcal{U}(t)$). Lower objective value and outside MSB goals are desired. The results show that our proposed algorithms outperform the other algorithms in the sense that (a) (b) they result in the lowest objective value across all the percentile, and (c) their average remains the lowest within an episode. }
    \label{fig:overall_performance}
\end{figure*}

\begin{figure*}[htbp]
    \centering
    \subfloat[Server Movement Cost]{\label{fig:eps_eq_1} \includegraphics[scale=0.37]{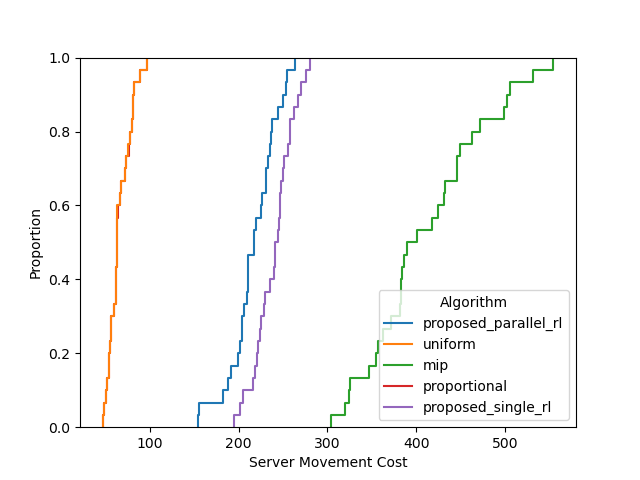}}
    \subfloat[Outside Rack Spread Goal]{\label{fig:eps_eq_2} \includegraphics[scale=0.37]{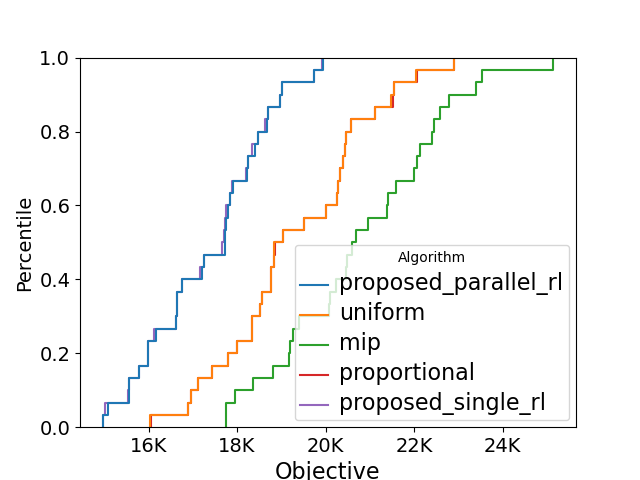}}
    \subfloat[Largest MSB]{\label{fig:eps_eq_4} \includegraphics[scale=0.37]{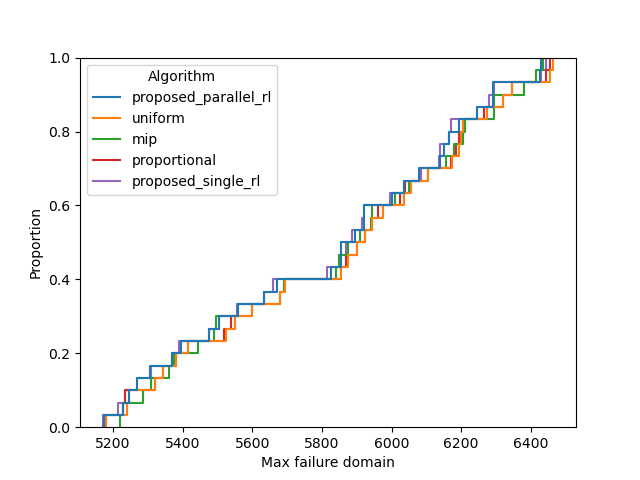}}
    \caption{The empirical CDF comparison concludes that (a) the proportional and uniform baselines induce the least server movement cost, (b) our proposed algorithms use the least resource outside rack spread goal, and (c) all the algorithms perform equally well in maintaining the size of the largest MSB.}
    \label{fig:spread}
\end{figure*}

\begin{figure*}[htbp]
    \centering
    \subfloat[Capacity Redundancy]{\label{fig:eps_eq_6} \includegraphics[scale=0.37]{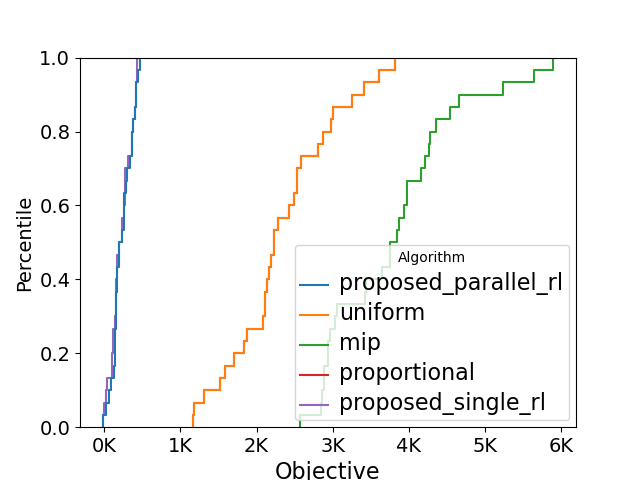}}
    \subfloat[Capacity Redundancy Variation]{\label{fig:eq_6} \includegraphics[scale=0.37]{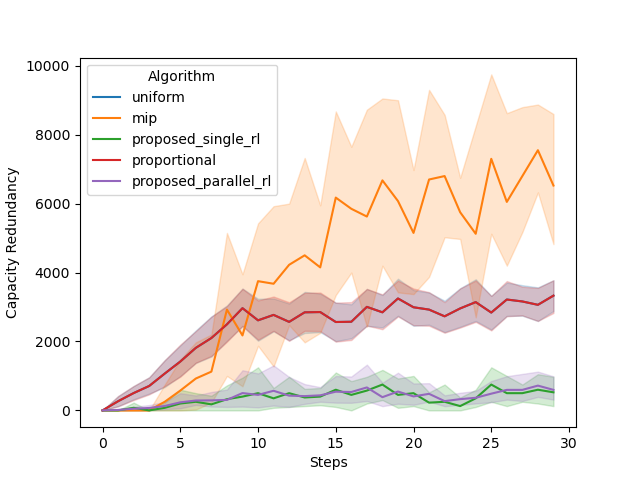}}
    \subfloat[Capacity Guarantee Violation]{\label{fig:eq_6_violation} \includegraphics[scale=0.37]{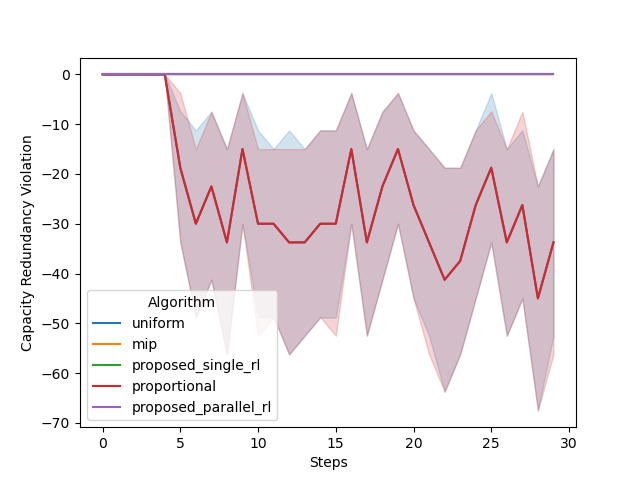}}
    \caption{We conclude that the resource utilization of our proposed algorithms is the most efficient and reliable by (a) their capacity redundancy is the lowest across the percentile, (b) their average capacity redundancy is the lowest for all time steps in an episode, and (c) they do not violate the redundancy constraint.}
    \label{fig:utilization}
\end{figure*}

We compare the performance of our proposed algorithms with the baseline algorithms in this section. In Figures \ref{fig:overall_performance}, \ref{fig:spread}, and \ref{fig:utilization}, we compare the baselines and our algorithm on the overall performance, resulting in server movement cost, spread, size of the largest fault domain, and induced capacity redundancy. 
We tested the performance as the system scale increases in Table \ref{table: 4}. Finally, we discuss how the decision tree model explains the decision-making of the PPO agent in Figure \ref{fig: DT}.
Simulation results such as the convergent rate of our proposed algorithm and the relationship between the terms in the objective function are presented in Figure \ref{fig:conv} in Appendix \ref{apdx:train} and Figure \ref{fig:varying_server_cost} in Appendix \ref{apdx:extra}, respectively.



In Figures \ref{fig:overall_performance}, \ref{fig:spread}, and \ref{fig:utilization}, we compare the overall performance, key metrics, and the resource utilization of the algorithms based on $30$ episodes of the performance data. Two kinds of plots are used to show the performance difference. The first kind compares the empirical cumulative distribution function (CDF) of a metric, where the horizontal axis represents the metric value and the vertical axis represents the percentile. The second kind compares the performance induced at each time step within an episode, where the horizontal axis represents the time steps and the vertical axis represents the metric values.

In Figure \ref{fig:obj_cdf} and \ref{fig:obj_cdf_vanilla}, we compare the empirical CDF of the resulting objective value. Because the complete PPO, complete TD3, and the random baseline perform much worse than the other algorithms, we divide the comparison into two figures for better comparison. A lower objective value is desired as it indicates better performance. The horizontal axis represents the objective value accumulated in an episode, while the vertical axis represents the percentile of an episode's accumulated objective value among all the episodes' objective values generated by a specific algorithm. The overall performance of the two proposed algorithms is very close, indicating that using a single agent can achieve similar performance to using parallel agents. The uniform and proportional baselines show similar performance. This can be attributed to the fact that the initial outputs of the proportional baseline are always uniform. The slight differences in their outputs may not significantly affect the generated server-to-reservation mappings by the action converter. The proposed algorithms consistently achieve the lowest objective values across all percentiles, outperforming the baselines. This demonstrates the effectiveness of the proposed algorithms in optimizing the objective function. The medians of our proposed algorithms, the uniform and proportional baselines, and the MIP solver are $38000$, $42000$, and $44000$, respectively, to name a few. The objective value obtained by complete PPO and complete TD3 range from $90000$ to $130000$, which indicates that the training of the complete PPO and complete TD3 agents converges at a suboptimal point and verifies that training an RL agent directly for a large-scale system is impractical. Since the complete PPO, complete TD3, and the random baseline perform much worse than the others, we omit their performance comparison with the proposed algorithms for the following content.

In Figure \ref{fig:obj_step}, the variation of the objective value within an episode is depicted, providing a different perspective for comparing performance. The horizontal axis represents the time steps within an episode, while the vertical axis displays the corresponding objective value. As the time steps within an episode increase, the performance difference between the proposed algorithms and the baseline algorithms becomes more pronounced. This suggests that the proposed algorithms exhibit better optimization capabilities and consistently improve the objective value over time compared to the baseline algorithms.
Figure \ref{fig:obj_step} supports the conclusion made in Figure \ref{fig:obj_cdf}, reinforcing the finding that the proposed algorithms outperform the baselines in terms of objective value optimization.
We can conclude that our proposed algorithm performs better than the proportional baseline and the MIP by $10\%$ and $15\%$, respectively, at the median of the experiments. 

Figure \ref{fig:eps_eq_1} illustrates that both the uniform and proportional baselines result in the lowest server movement cost. Additionally, the two proposed algorithms outperform the MIP solver significantly in terms of server movement cost. The results of the uniform and proportional baselines imply that uniformly getting resources for all the scopes renders the least server movement cost. However, the objective function of the problem is primarily influenced by the resource spread components as demonstrated in Figures \ref{fig:eps_eq_2} and \ref{fig:overall_performance}. Figure \ref{fig:eps_eq_4} shows that all the algorithms maintain the largest failure domain equally well. It implies that these algorithms are comparable in terms of their ability to handle failures effectively.

In Figures \ref{fig:eps_eq_6} and \ref{fig:eq_6}, we compare the capacity redundancy discussed in Eq. \eqref{eq: capacity_redundancy}. The capacity redundancy indicated on the horizontal axis of Figure \ref{fig:eps_eq_6} and the vertical axis of Figure \ref{fig:eq_6} represents the average of $g_{2,l,e}(t)$ for all reservations and server types.
It's evident that our proposed algorithms exhibit the most efficient use of additional resources to ensure a sufficient supply when confronted with the largest MSB failures because the induced capacity redundancy is the least among all the algorithms.
Besides the overall induced capacity redundancy, we also show the capacity guarantee violation in Figure \ref{fig:eq_6_violation}, which tallies the total deficit. Figure \ref{fig:eq_6_violation} shows that the proportional and uniform baselines violate the capacity redundancy constraint starting at step $5$ till the end. It's important to note that we designed the outputs of the uniform and proportional methods to meet the capacity redundancy constraint initially in Section \ref{subsec: baselines}. This violation may be attributed to server competition when demand surges. In our low-level MILP, server requests that cannot be fulfilled due to rack capacity limits are redistributed to other racks, and the low-level MILP does not ensure constraint satisfaction during redistribution. Moreover, it's noteworthy that our proposed agent demonstrates the ability to learn and avoid constraint violations, as evident from Figure \ref{fig:eq_6_violation}, where no violations occur for our proposed algorithm.

Table \ref{table: 4} compares the scalability of the proposed single agent algorithm with the MIP solver based on testing time and training time, measured on the same machine with $32$ CPU cores and $64$ GB RAM. The table includes results for varying numbers of servers, ranging from $500$ to $10,000$. Testing time represents the time required to finish one episode, i.e., $30$ consecutive decision-makings. On the other hand, the training time is the expected time for training our RL agent for $1000$ episodes. The testing time in Table \ref{table: 4} are averaged over $30$ episodes, while the training time represents the time required to train the agent for $1000$ episodes. Our proposed algorithm shows manageable training times even with a system containing $10000$ servers. Moreover, the proposed algorithm exhibits significantly lower testing times compared to the MIP solver, highlighting greater flexibility in system management. Though the training time may take hours or even days for a very large system, most of the training needs can be anticipated and planned in advance. Moreover, updating the RL model is simple and convenient. Besides the training and testing time, our proposed algorithm also outperforms the MIP as indicated by both the table and the previous figures. Note that the training time for our proposed parallel agents algorithm can be further reduced by training each model on different devices simultaneously. The training time of the proposed parallel agents can be $14$, $36$, $56$ minutes for systems with $500$, $1000$, and $10000$ servers, respectively.

\begin{table}[htbp] 
\small
\caption{Summary of the algorithm scalability} \label{table: 4}
\centering
\resizebox{\columnwidth}{!}{%
 \begin{tabular}{| c || c | c | c | c | c |} 
 \hline
 (Algorithm, $|B|$) & (MIP, 500) & (MIP, 1000) & (Proposed, 500) & (Proposed, 1000) & (Proposed, 10000) \\
 \hline\hline
Testing time (sec) & 2329  & 3510 & 10  & 26 & 39\\
 \hline
 \makecell[c]{Training time (min)} & N/A & N/A &53  & 142 & 224 \\
 \hline
 \makecell[c]{Performance $\mathcal{U}(t)$} &25746 &44213 &21128 &37795  &1221747  \\
 \hline
 \end{tabular}
 }
\end{table}



\section{Conclusion and Future Works} \label{sec: conclusion}

This work considers optimizing server-to-reservation mapping for a dynamic RAS constituting datacenters, MSBs, racks, and heterogeneous servers.  A DRL agent and action converter determine resource supply at the MSB level, followed by a MILP for server-to-reservation mapping. By decoupling the problem by server type, independent optimizations are performed. Two algorithm variants, single and parallel DRL agents, are compared with heuristic and MIP-based methods, showing superior performance and reduced computation time. 

Future steps for this work include extending the optimization problem to incorporate practical failure arrivals and addressing containers' placement on servers to satisfy specific service requirements. These extensions would further enhance the practical applicability and effectiveness of the proposed algorithms in real-world scenarios.

\appendices






\section{Related Works}

\subsection{Cluster manager scheduling and solvers}\label{subsec: non_drl_approaches}

Server resource allocation problems are often solved as mathematical optimization problems \cite{abuzaid2021contracting,xiang2015joint,gog2016firmament,kumar2015bwe,stefanello2019hybrid,li2020traffic,newell_ras_2021}. These problems involve determining the allocation of available resources, such as accelerators, servers, or network links, to clients, such as jobs, data shards, or traffic commodities, based on their specific needs. However, these mathematical optimization methods are frequently computationally expensive \cite{narayanan2021solving}. 
Moreover, integer-linear programs for resource allocation problems can be even more computationally expensive \cite{lee2015efficient}, particularly in large-scale systems where the number of variables can reach millions (e.g., considering one variable for each client-resource pair). Consequently, the solution times for these problems can be long for solvers to generate a response (as a reference, SCS \cite{scs} requires 8 minutes to allocate a cluster with 1000 jobs \cite{o2016conic}). 
Additionally, it is highly desirable for the system to make decisions considering the future, while classical stochastic optimization methods developed by the math programming community are usually prohibitively more expensive than deterministic problems \cite{shapiro2007tutorial}.
Many existing production solutions, such as the Gavel job scheduler \cite{narayanan2020heterogeneity}, the Accordion load balancer \cite{serafini2014accordion} for distributed databases, and BwE \cite{kumar2015bwe} for traffic engineering, can experience performance bottlenecks with an increasing number of clients and resources to allocate. Consequently, directly solving optimization problems often becomes impractical \cite{narayanan2021solving}, leading production systems to frequently employ heuristic or meta-heuristic solutions (as in \cite{stefanello2019hybrid}) that are cheaper and faster to compute for such problems. However, many heuristic solutions are specifically designed for particular problems like job scheduling or bandwidth allocation, which may limit their applicability to our problem. Additionally, most existing algorithms for cloud computing systems are primarily focused on workload assignment to hardware resources. To the best of our knowledge, RAS \cite{newell_ras_2021} and this work are the only two approaches addressing the server clustering problem while considering failure occurrence. Nevertheless, both heuristic approaches and MIP solvers may not be optimal for dynamic systems.

\subsection{Deep Learning and Reinforcement Learning on Cluster Management} \label{subsec: rl_related_works}
The time-varying nature and complex architecture of cloud systems make resource allocation problems challenging. Deep Reinforcement Learning (DRL) has demonstrated success in addressing dynamic and complex systems, including robotics \cite{8675643, chen2022hierarchical, chen2023multi}, autonomous vehicle coordination \cite{hu2019interaction, al2019deeppool, chen2021deepfreight}, communication and network packet delivery \cite{ye2015multi,geng2023reinforcement, gonzalez2023asap}, including works through multi-agent reinforcement learning \cite{mondal2022can, pac,gadiraju2023optimization}. Consequently, DRL has also been adopted to solve resource allocation problems in cloud systems. In the context of cloud systems, RL has been applied to design congestion control protocols \cite{winstein2013tcp} and develop simple resource management systems by treating the problem as learning packing tasks with multiple resource demands \cite{mao2016resource}. Luan et al. \cite{luan2019sched2} proposed a GPU cluster scheduler that leverages DRL for intelligent locality-aware scheduling of deep learning training jobs. The authors of \cite{li2019deepjs} applied RL to minimize the makespan of a job set.  \cite{mao2019learning} proposed the use of actor-critic methods with input-dependent baselines for distributing computation clusters. The authors of \cite{zhang2020learning} proposed a deep Q-learning approach to solve the CPU-GPU heterogeneous computing scheduling problem based on the running state and task characteristics of the cluster environment. 

\subsection{Capacity Reservation}
A traditional approach to improving resource allocation efficiency is to group servers into physical clusters, which reduces the number of candidate servers for container placement \cite{mao2016resource}. However, this method often leads to uneven utilization of clusters, suboptimal server allocation, and challenges in recovering from data center failures, causing difficulties for service owners \cite{swarup2021task}. 
To address these issues, the authors of \cite{tang2020twine} introduced a novel approach, Twine, by organizing servers into reservations and creating a mega server pool as a backup for each reservation. This design allows for flexible system management and eliminates the problem of stranded resources in small physical clusters. Building upon Twine, the Resource Allowance System (RAS) introduced in \cite{newell_ras_2021} further enhances resource allocation efficiency. RAS separates the server-to-reservation assignment from container placement, allowing for more optimized resource allocation. It formulates the server assignment problem, considering factors such as hardware failures at the scope of racks and MSBs, data center maintenance, and heterogeneous hardware. The problem is then periodically solved using a Mixed Integer Programming (MIP) solver. 

Instead of directly focusing on optimizing container placement, as was the case in prior works outlined in Appendix \ref{subsec: non_drl_approaches} and \ref{subsec: rl_related_works}, this study emphasizes the attainment of long-term optimization for server-to-reservation assignments. Existing approaches can subsequently handle the container placement for each reservation. It's important to note that our approach enhances the efficiency of container-to-server placement by refining the available server options for each container to those that are more likely to fulfill the specified requirements. Additionally, our optimization process incorporates considerations for failure occurrences to ensure capacity guarantees.
Since system management involves a series of decision makings, optimizing the system considering only the current system status could not achieve long-term optimum. Therefore, we proposed a two-level DRL-based algorithm where a DRL agent and our proposed action converter allocate resources for each reservation at the MSB level, and a low-level MILP solver solves the server-to-reservation mapping based on the results from the first level.
\newpage
\section{Notation summary} \label{apdx: notation_summary}
\begin{table}[tbhp] 
\caption{Summary of Optimization Problem Parameters} \label{table: op_param_summary}
\centering
 \begin{tabular}{||c|| c ||} 
 \hline
 Notation &Description  \\ [0.5ex] 
 \hline\hline
 $D$ &Set of datacenters \\ 
 \hline
 $F$ &Set of MSBs \\ 
 \hline
 $K$ &Set of racks \\ 
 \hline
 $B$ &Set of servers \\ 
 \hline
 $L$ &Set of all reservations \\
 \hline
 $E$ &Set of server types \\ 
 \hline
 $x_{b,l}(t)$ &\makecell{Assignment variable which is 1 if server $b$\\ is assigned to reservation $l$ and 0 otherwise} \\ 
 \hline
 $\alpha_b(t)$ &\makecell{The reservation that server $b$ is\\ assigned to at time slot $t$}\\
 \hline
 $M_b$ &Movement cost of server $b$\\
 \hline
 $\kappa$ &Cost of the maximum MSB failure\\
 \hline
 $\beta$ &Cost of resource allocation outside of spread goals\\
 \hline
 $\alpha^{K, F}$ &\makecell{Proportional limit of reservation of\\ spread in $K$ (rack) or $F$ (MSB fault domain)}\\
 \hline
 $U_{b}$ &RRU of server $b$\\
 \hline
 $\mathbf{C}_{l}(t)$ &Capacity desired for reservation $l$\\
 \hline
 $\Psi^{K,F,D}$ &\makecell{Partition of servers based on $K$ \\(rack), $D$ (datacenter), or $F$ (MSB fault domain)}\\
 \hline
 $\Psi^{K}_k$, $\Psi^{F}_f$, $\Psi^{D}_d$ &Servers in rack $k$, in MSB $f$, or in datacenter $d$\\
 \hline
 $\Phi^{F}_f$ &Racks in MSB $F$ (MSB fault domain)\\
 \hline
 $A_{d,l,e}$ &\makecell{Affinity of reservation $l$ to a datacenter\\ regarding server type $e$}\\
[1ex] 
 \hline
 \end{tabular}
\end{table}

\section{Algorithms} \label{apdx: algorithms}
\begin{algorithm}[bhtp]
  \DontPrintSemicolon
  \SetAlgoLined
  \KwIn{$r$, $\{n_{l,e,f}(t), \forall l\in L, \forall f\in F\}$}
  \KwOut{$\{m_{l,e,k}(t), \forall l\in L, \forall k\in K\}$ --- the number of servers reservation $l$ should take from the racks.}
  \For{$f$ = 1 \KwTo $|F|$}{
  Sort the racks in the MSB $f$ according to the number of servers previously assigned to reservation $l$ in the descending order\;
  $p \gets $ quotient of $n_{l,e,f}(t)$ by the number of racks in MSB $f$\;
  $q \gets $ modulus of $n_{l,e,f}(t)$ by the number of racks in MSB $f$\;
  $counter \gets 0$\;
  \For{$k$ in $\Phi^F_f$}
  {
    \eIf{$counter < q$}
    {
      $m_{l,e,k}(t) \gets p+1$\;
    }{
      $m_{l,e,k}(t) \gets p$\;
    }
    $counter \gets counter+1$\;
  }
  }
  \caption{Get the number of servers to take from the racks in MSBs for reservations}\label{alg: even_servers_from_racks}
\end{algorithm}
This section shows the details of our proposed low-level MILP solver which first determines the number of server requests for each rack by Algorithm \ref{alg: even_servers_from_racks} and then derives the server-to-reservation mapping by Algorithm \ref{alg: actual_mapping}, and presents how the PPO agent is trained under our framework in Algorithm \ref{alg: ppo_train}.
\begin{algorithm}[h!tbp]
  \DontPrintSemicolon
  \SetAlgoLined
  \KwIn{$\{m_{l,e,k}(t), \forall l\in L, \forall k\in K\}$}
  \KwOut{$\{x_{b,l}(t), \forall b\in B_e, \forall l\in L\}$}
  \For{$k$ = 1 \KwTo $|K|$}{
    $u_{k}(t) \gets \sum_{l=1}^{|L|}m_{l,k}(t)-|\Psi^K_k|$\;
  }
  \For{$f$ = 1 \KwTo $|F|$}{
      \For{$k$ in $\Phi^F_f$}{
            $i \gets |R|$\;
            \While{$u_k(t)>0$ and $i\geq0$}
            {
                \eIf{$m_{i,k}(t)<=u_k(t)$}{
                    $u_k(t) \gets u_k(t) - m_{i,k}(t)$\;
                    $m_{i,k}(t) \gets 0$\;
                }{
                    $u_k(t) \gets 0$\;
                    $m_{i,k}(t) \gets m_{i,k}(t)-u_k(t)$\;
                }
                Try to first move $\min(u_k(t), m_{i,k}(t))$ requests to racks in $\Phi^F_f$\;
                If it is not possible, move them to the racks in the other MSBs.\;
                $i \gets i-1$\;
            }
      }
  }
  \For{$k$ = 1 \KwTo $K$}{
    \For{$b$ in $\Psi^K_k$}{
        \eIf{$\alpha_b(t-1) == 1$ and $m_{\alpha_b(t-1), k}>0$}{
            $m_{\alpha_b(t-1), k} \gets m_{\alpha_b(t-1), k}-1$\;
            $x_{b, \alpha_b(t-1)}(t) \gets 1$\;
        }{}
    }
  }
  \caption{Pseudocode of Algorithm \ref{alg: actual_mapping}} \label{alg: actual_mapping_details}
\end{algorithm}

\begin{algorithm}[h!tbp]
\DontPrintSemicolon
\SetAlgoLined
Initialize the size of the memory, learning rate, etc.\;
Initialize one agent or $|E|$ agents\;
\For{$e=1$ \KwTo $|E|$}{
Start training the $e-th$ agent if this is for the parallel agents case\;
\For{$episode=1$ \KwTo $\Gamma$}{
Initialize the capacity requests, reservations, and servers\;
\For{$t=1$ \KwTo $\mathcal{T}$}{
\For{$l=1$ \KwTo $|L|$}{
    Combine the information of the environment and actions of reservations $1,...,l-1$, into $\mathbf{s}_{l,e}(t)$\;
    Send $\mathbf{s}_{l,e}(t)$ to the PPO agent and get $\mathbf{a}_{l,e}(t)$\;
    Convert $\mathbf{a}_{l,e}(t)$ to numbers of servers to take by reservation $l$, $\{n_{l,e,f}(t), \forall f\in F\}$, by \eqref{eq: num_servers_get_msb}\;
    Get numbers of servers to take from the racks, $\{m_{l,e,k}(t), \forall l\in L, \forall k\in K\}$, by Algorithm \ref{alg: even_servers_from_racks}\;
}
Get server-to-reservation mapping, $\{x_{b,l}(t)\}$ for all $b$ and $l$ by Algorithm \ref{alg: actual_mapping}\;
\For{$l=1$\KwTo$|L|$}{
    Calculate the reward of reservation $l$ by \eqref{eq: reward_reservation}\;
    Store the state, action, and reward in the memory of the PPO agent\;
}
Update the environment with the server-to-reservation mapping, $\{x_{b,l}(t), \forall b\in B, \forall l\in L\}$\;
$t \gets t+1$\;
}
}
}
\caption{Training the PPO agent \label{alg: ppo_train}}
\end{algorithm}

\section{Training Results}\label{apdx:train}
\begin{figure*}[htbp]
    \centering
    \subfloat[][]{\label{fig:conv_reward} \includegraphics[scale=0.4]{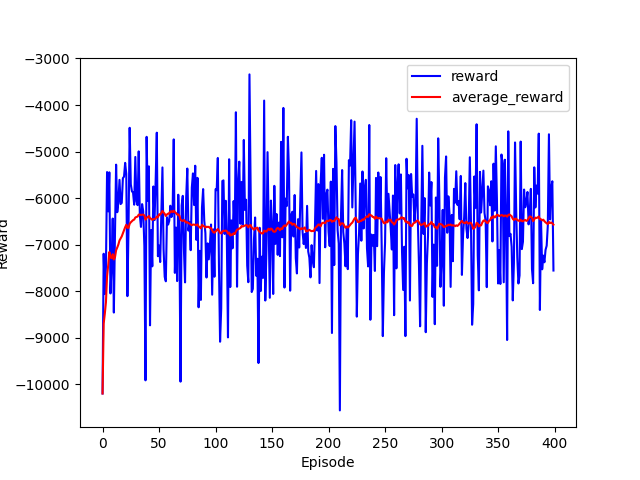}}
    \subfloat[][]{\label{fig:conv_obj} \includegraphics[scale=0.4]{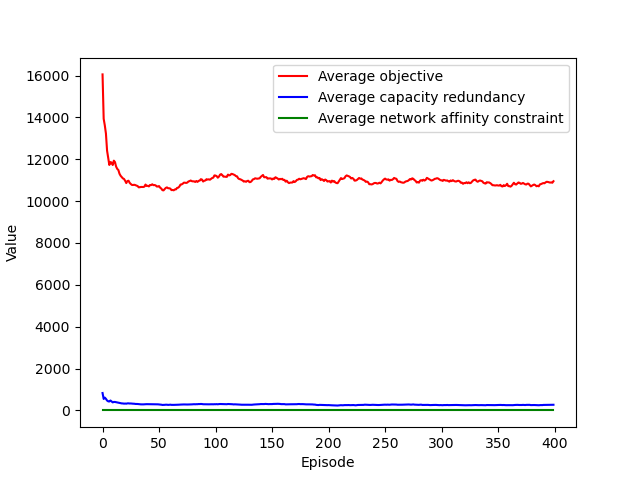}}
    \caption{Algorithm convergence simulations for the server type $1$. Both our proposed parallel agents and our proposed single-agent algorithms show similar convergence patterns. The result shows that the moving average of the reward, the objective value, the capacity redundancy constraint, and the network affinity constraint of our proposed algorithms converge after $100$ episodes.}
    \label{fig:conv}
\end{figure*}

We first show the convergence rate in terms of the number of episodes required to reach convergence. In Figures \ref{fig:conv_reward} and \ref{fig:conv_obj}, we train the agent for server type $1$ for $400$ episodes. Figure \ref{fig:conv_reward} shows the variation of the reward and its moving average. We observe that the moving average converges after $100$ episodes. Similarly, in Figure \ref{fig:conv_obj}, we observe that there is little variation of the moving average of the objective value and constraint $6$ after $100$ episodes. 
Training plots for the rest of the server types are omitted for the sake of repetition. 

\section{Extra Simulation Results}\label{apdx:extra}

\begin{figure}[htbp]
    \centering
    \subfloat[Server Movements and Capacity Redundancy]{\label{fig:varying_server_cost_1_6} \includegraphics[width=2.8in, height=1.8in]{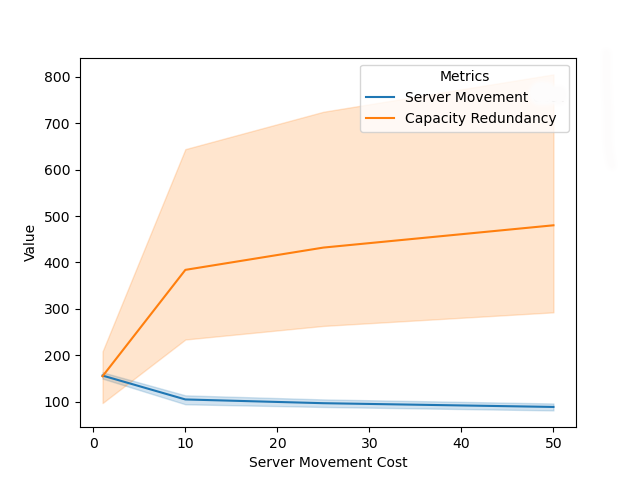}} \\
    \subfloat[Outside Resource Spread Goals]{\label{fig:varying_server_cost_2_3_4} \includegraphics[width=2.8in, height=1.8in]{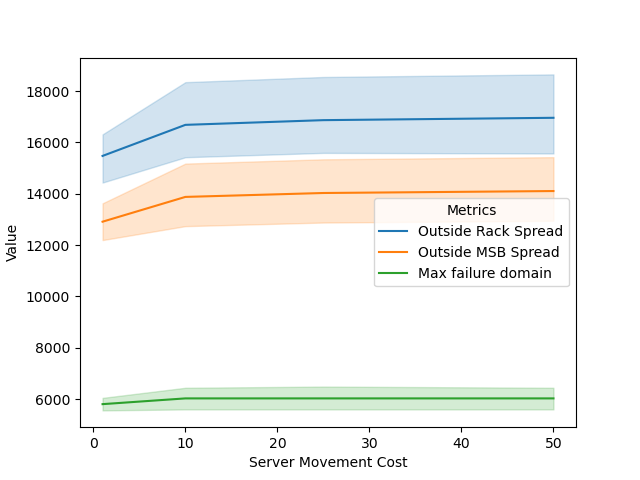}}
    \caption{The increase of the unit server movement cost helps further minimize the server movements, but induces higher capacity redundancy and worse resource spread.}
    \label{fig:varying_server_cost}
\end{figure}

In Figure \ref{fig:varying_server_cost}, we examine the trade-off between metrics in the optimization problem as the server movement cost ($M_b$) increases. 
Since the single agent algorithm and parallel agents algorithm have close performance, we test only the performance of the single agent. 
We plot Figure \ref{fig:varying_server_cost} based on $5$ episodes of performance data obtained at different server movement costs: $5$, $10$, $25$, and $50$. 
Figure \ref{fig:varying_server_cost_1_6} indicates that the total number of server movements decreases as the server movement cost increases. We also observe slight rise on the resource spread and capacity redundancy as shown in Figures \ref{fig:varying_server_cost_1_6} and \ref{fig:varying_server_cost_2_3_4}. This increase might be due to the utilization of additional servers to satisfy incoming demands and avoid the high server movement cost.

\section{Verification by Decision Tree Models} \label{subsec: dt}
Though reinforcement learning has been successfully applied to various sequential decision-making problems, the difficulty of predicting and verifying the behavior of the RL agents often hinders their real-world implementation. Therefore, we utilize decision tree models to verify the behavior of our DRL-based algorithm. Specifically, we collect the state and action pairs, $\mathbf{a}_{l,e}(t)$ and $\mathbf{s}_{l,e}(t)$, by running our trained DRL-based algorithm. We focus on inspecting the total number of servers the agent decides to allocate given a state instead of the distribution of allocated servers among the MSBs by decision tree models because the server distribution among MSBs determined by the low-level MILP solver is usually uniform. Then, we group the state and action pairs into multiple dataset by the capacity demand $C_{l,e}(t)$ and train a decision tree model for each dataset. The structure of the decision tree model makes it easy to get a plausible explanation of how the state values lead to a decision and identify unreasonable extreme cases, so a domain expert can review and find out unreasonable decision-makings and further improve the RL agent. In this paper, we add filters in the low-level MILP solver to correct the extreme cases that happens once in tens of thousands of state-action pairs.

\begin{figure*}[htbp]
    \centering
    \includegraphics[width=\textwidth]{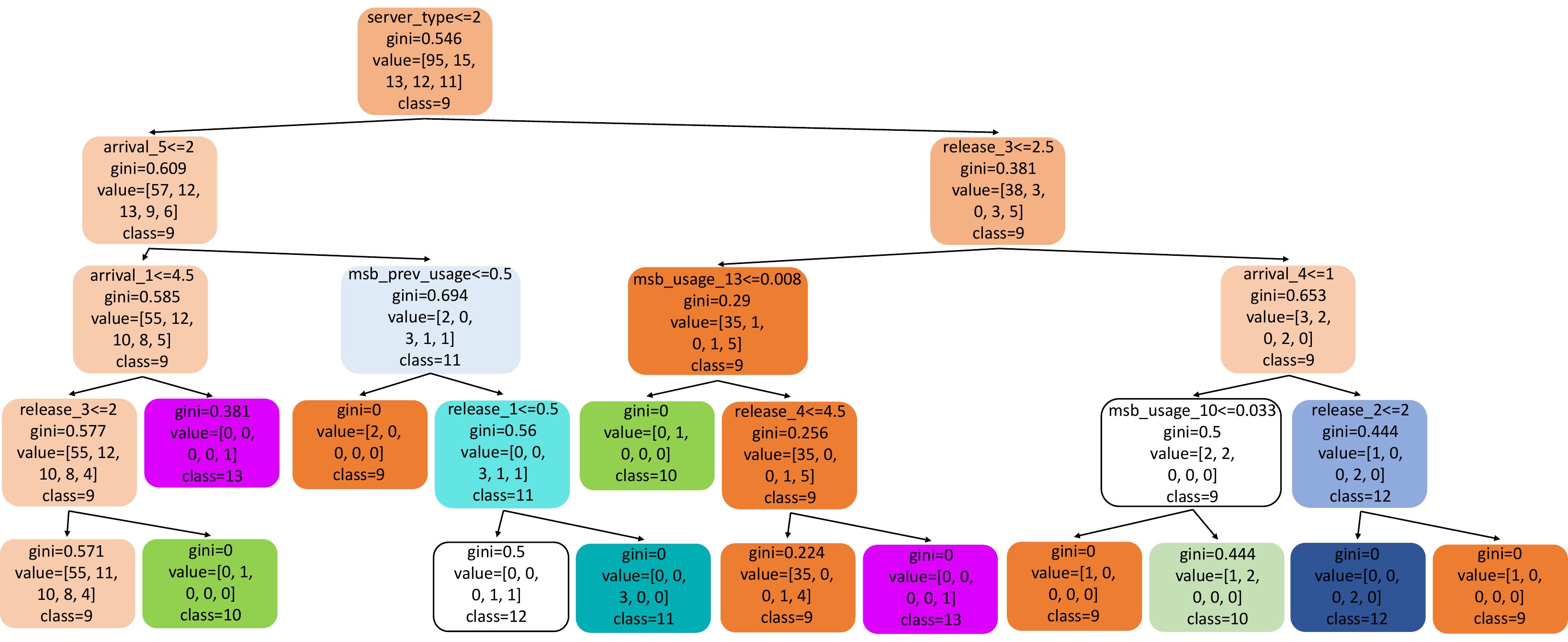}
    \caption{A decision tree for the case of the capacity demand being equal to $8$ servers.}
    \label{fig: DT}
\end{figure*}

In Figure \ref{fig: DT}, we have a four-level decision tree trained using state-action pairs where the capacity demand in the state is fixed at $8$ servers. The decision tree takes the state variable as input and predicts the number of servers to allocate as output.
Each non-leaf node in the decision tree is responsible for classifying data samples based on one of the state variables selected to minimize the Gini impurity metric. The first line in a non-leaf node consists of a conditional statement regarding one of the state variables. Data samples adhering to the condition are directed to the left branch, while those that do not meet the condition go to the right branch. The state variables, such as `arrival\_k' and `release\_k', signify the predicted demand increase or decrease in the future k-th time slot, respectively. Conversely, state variables like `msb\_usage\_f' and `msb\_prev\_usage\_f' represent the resource usage of a reservation on the f-th MSB during the current and previous time slots, respectively.
The value array present in each node denotes the number of samples falling into each class. The class value at the bottom line of each node represents the dominant decision made by the agent among the data samples within that node. For example, the value array of the root node indicates that there are 95 samples where the agent decides to supply 9 servers, 15 samples for supplying 10 servers, 13 samples for supplying 11 servers, 12 samples for supplying 12 servers, and 11 samples for supplying 13 servers to the reservation, respectively. Thus, the class value of the root node is 9.
By examining Figure \ref{fig: DT}, we can discern that the agent frequently allocates $9$ servers, especially when the anticipated future demand arrivals and expirations are modest. For instance, consider the paths `server\_type$\leq$2'--`arrival\_5$\leq$2'--`arrival\_1$\leq$4.5'--`release\_3$\leq$2' and `server\_type$>$2'--`release\_3$\leq$2.5'--`msb\_usage\_13$\leq$0.008'--`arrival\_4$\leq$4.5'.
It's worth noting that if we increase the levels of the decision tree, we can gain a more detailed understanding of the agent's behavior from the decision tree.

\section{Evaluation on Large-Scale System}
\begin{figure}
    \centering
    \includegraphics[width=0.45\textwidth]{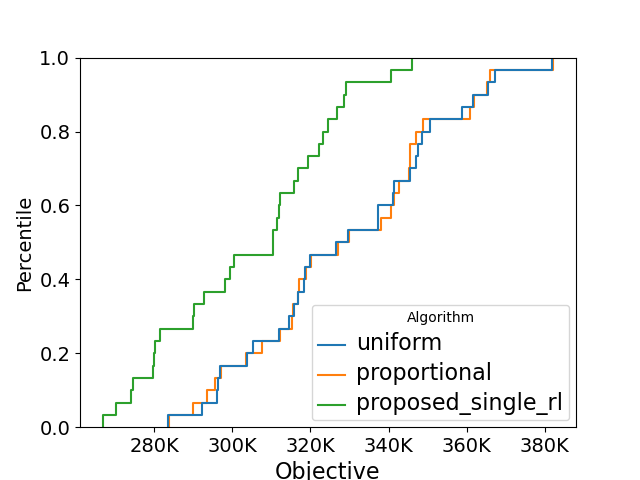}
    \caption{Objective value comparison for a large scale system.}
    \label{fig:large}
\end{figure}

In Figure \ref{fig:large}, we test the performance of the proposed algorithm for a large scale system with $5$ data centers, $50$ MSBs, $500$ racks, $100,000$ servers, and $100$ reservations. We also make the following modifications to the simulator to better resemble real world cloud computing system:
\begin{enumerate}
    \item Similar to the original setting, the capacity requests contribute to the demand of the corresponding reservation upon arrival, but they do not expire. The difference is that users of the reservation have the ability to send resource unsubscribing requests to reduce the reservation's total demand. The frequency of occurrence of capacity requests and unsubscribing requests adheres to real data center statistics.
    \item Servers within a rack are of the same server type.
    \item The number of reservations and servers changes with time. 
\end{enumerate}
Due to the impracticality of employing MIP and pure RL algorithms for such a large system, we compare our algorithm with the proportional and uniform baselines.

Figure \ref{fig:large} illustrates the comparison between the proposed algorithm and the uniform and proportional baselines. The results indicate that the proposed algorithm outperforms the two baselines by approximately 10\%. This underscores the importance of the agent's ability to adjust the weight of resources from the MSBs in a timely manner, especially in system settings where uniform allocation may be considered optimal. Furthermore, these findings demonstrate that our algorithm effectively addresses systems that closely resemble real-world data centers.

\bibliographystyle{IEEEtran}
\bibliography{reference}

\begin{thebibliography}{10}
\providecommand{\url}[1]{#1}
\csname url@samestyle\endcsname
\providecommand{\newblock}{\relax}
\providecommand{\bibinfo}[2]{#2}
\providecommand{\BIBentrySTDinterwordspacing}{\spaceskip=0pt\relax}
\providecommand{\BIBentryALTinterwordstretchfactor}{4}
\providecommand{\BIBentryALTinterwordspacing}{\spaceskip=\fontdimen2\font plus
\BIBentryALTinterwordstretchfactor\fontdimen3\font minus \fontdimen4\font\relax}
\providecommand{\BIBforeignlanguage}[2]{{%
\expandafter\ifx\csname l@#1\endcsname\relax
\typeout{** WARNING: IEEEtran.bst: No hyphenation pattern has been}%
\typeout{** loaded for the language `#1'. Using the pattern for}%
\typeout{** the default language instead.}%
\else
\language=\csname l@#1\endcsname
\fi
#2}}
\providecommand{\BIBdecl}{\relax}
\BIBdecl

\bibitem{Google_Verma2015}
A.~Verma, L.~Pedrosa, M.~R. Korupolu, D.~Oppenheimer, E.~Tune, and J.~Wilkes, ``Large-scale cluster management at {Google} with {Borg},'' in \emph{Proceedings of the European Conference on Computer Systems (EuroSys)}, Bordeaux, France, 2015.

\bibitem{Hindman2011}
\BIBentryALTinterwordspacing
B.~Hindman, A.~Konwinski, M.~Zaharia, A.~Ghodsi, A.~D. Joseph, R.~Katz, S.~Shenker, and I.~Stoica, ``Mesos: A platform for {Fine-Grained} resource sharing in the data center,'' in \emph{8th USENIX Symposium on Networked Systems Design and Implementation (NSDI 11)}.\hskip 1em plus 0.5em minus 0.4em\relax Boston, MA: USENIX Association, Mar. 2011. [Online]. Available: \url{https://www.usenix.org/conference/nsdi11/mesos-platform-fine-grained-resource-sharing-data-center}
\BIBentrySTDinterwordspacing

\bibitem{tang2020twine}
C.~Tang, K.~Yu, K.~Veeraraghavan, J.~Kaldor, S.~Michelson, T.~Kooburat, A.~Anbudurai, M.~Clark, K.~Gogia, L.~Cheng \emph{et~al.}, ``Twine: A unified cluster management system for shared infrastructure,'' in \emph{Proceedings of the 14th USENIX Conference on Operating Systems Design and Implementation}, 2020, pp. 787--803.

\bibitem{newell_ras_2021}
A.~Newell, D.~Skarlatos, J.~Fan, P.~Kumar, M.~Khutornenko, M.~Pundir, Y.~Zhang, M.~Zhang, Y.~Liu, L.~Le \emph{et~al.}, ``Ras: continuously optimized region-wide datacenter resource allocation,'' in \emph{Proceedings of the ACM SIGOPS 28th Symposium on Operating Systems Principles}, 2021, pp. 505--520.

\bibitem{Verbitski2017}
A.~Verbitski, A.~Gupta, D.~Saha, M.~Brahmadesam, K.~Gupta, R.~Mittal, S.~Krishnamurthy, S.~Maurice, T.~Kharatishvili, and X.~Bao, ``Amazon aurora: Design considerations for high throughput cloud-native relational databases,'' in \emph{SIGMOD 2017}, 2017.

\bibitem{aws}
\BIBentryALTinterwordspacing
AWS. (2023) On-demand capacity reservations. [Online]. Available: \url{https://docs.aws.amazon.com/AWSEC2/latest/UserGuide/ec2-capacity-reservations.html}
\BIBentrySTDinterwordspacing

\bibitem{google_web}
\BIBentryALTinterwordspacing
Google. (2023) Reservations of compute engine zonal resources. [Online]. Available: \url{https://cloud.google.com/compute/docs/instances/reservations-overview}
\BIBentrySTDinterwordspacing

\bibitem{mao2019learning}
H.~Mao, M.~Schwarzkopf, S.~B. Venkatakrishnan, Z.~Meng, and M.~Alizadeh, ``Learning scheduling algorithms for data processing clusters,'' in \emph{Proceedings of the ACM special interest group on data communication}, 2019, pp. 270--288.

\bibitem{DBLP:journals/corr/SchulmanWDRK17}
J.~Schulman, F.~Wolski, P.~Dhariwal, A.~Radford, and O.~Klimov, ``Proximal policy optimization algorithms,'' \emph{CoRR}, vol. abs/1707.06347, 2017.

\bibitem{curriculum_ronan}
\BIBentryALTinterwordspacing
Y.~Bengio, J.~Louradour, R.~Collobert, and J.~Weston, ``Curriculum learning,'' in \emph{Proceedings of the 26th Annual International Conference on Machine Learning}, ser. ICML '09.\hskip 1em plus 0.5em minus 0.4em\relax New York, NY, USA: Association for Computing Machinery, 2009, p. 41–48. [Online]. Available: \url{https://doi.org/10.1145/1553374.1553380}
\BIBentrySTDinterwordspacing

\bibitem{mondal2022can}
W.~U. Mondal, M.~Agarwal, V.~Aggarwal, and S.~V. Ukkusuri, ``On the approximation of cooperative heterogeneous multi-agent reinforcement learning (marl) using mean field control (mfc),'' \emph{Journal of Machine Learning Research}, vol.~23, no. 129, pp. 1--46, 2022.

\bibitem{hallak_contextual_2015}
A.~Hallak, D.~Di~Castro, and S.~Mannor, ``Contextual markov decision processes,'' \emph{arXiv preprint arXiv:1502.02259}, 2015.

\bibitem{python_mip}
\BIBentryALTinterwordspacing
H.~G.~S. Túlio A. M.~Toffolo. (2019) Python-mip. [Online]. Available: \url{https://www.python-mip.com}
\BIBentrySTDinterwordspacing

\bibitem{td3_fujimoto}
S.~Fujimoto, H.~van Hoof, and D.~Meger, ``Addressing function approximation error in actor-critic methods,'' 2018.

\bibitem{abuzaid2021contracting}
F.~Abuzaid, S.~Kandula, B.~Arzani, I.~Menache, M.~Zaharia, and P.~Bailis, ``Contracting wide-area network topologies to solve flow problems quickly.'' in \emph{NSDI}, 2021, pp. 175--200.

\bibitem{xiang2015joint}
Y.~Xiang, T.~Lan, V.~Aggarwal, and Y.-F.~R. Chen, ``Joint latency and cost optimization for erasure-coded data center storage,'' \emph{IEEE/ACM Transactions on Networking}, vol.~24, no.~4, pp. 2443--2457, 2015.

\bibitem{gog2016firmament}
I.~Gog, M.~Schwarzkopf, A.~Gleave, R.~N. Watson, and S.~Hand, ``Firmament: Fast, centralized cluster scheduling at scale,'' in \emph{12th USENIX Symposium on Operating Systems Design and Implementation (OSDI 16)}, 2016, pp. 99--115.

\bibitem{kumar2015bwe}
A.~Kumar, S.~Jain, U.~Naik, A.~Raghuraman, N.~Kasinadhuni, E.~C. Zermeno, C.~S. Gunn, J.~Ai, B.~Carlin, M.~Amarandei-Stavila \emph{et~al.}, ``Bwe: Flexible, hierarchical bandwidth allocation for wan distributed computing,'' in \emph{Proceedings of the 2015 ACM Conference on Special Interest Group on Data Communication}, 2015, pp. 1--14.

\bibitem{stefanello2019hybrid}
F.~Stefanello, V.~Aggarwal, L.~S. Buriol, and M.~G. Resende, ``Hybrid algorithms for placement of virtual machines across geo-separated data centers,'' \emph{Journal of Combinatorial Optimization}, vol.~38, pp. 748--793, 2019.

\bibitem{li2020traffic}
X.~Li and K.~L. Yeung, ``Traffic engineering in segment routing networks using milp,'' \emph{IEEE Transactions on Network and Service Management}, vol.~17, no.~3, pp. 1941--1953, 2020.

\bibitem{narayanan2021solving}
D.~Narayanan, F.~Kazhamiaka, F.~Abuzaid, P.~Kraft, A.~Agrawal, S.~Kandula, S.~Boyd, and M.~Zaharia, ``Solving large-scale granular resource allocation problems efficiently with pop,'' in \emph{Proceedings of the ACM SIGOPS 28th Symposium on Operating Systems Principles}, 2021, pp. 521--537.

\bibitem{lee2015efficient}
Y.~T. Lee and A.~Sidford, ``Efficient inverse maintenance and faster algorithms for linear programming,'' in \emph{2015 IEEE 56th Annual Symposium on Foundations of Computer Science}.\hskip 1em plus 0.5em minus 0.4em\relax IEEE, 2015, pp. 230--249.

\bibitem{scs}
B.~O'Donoghue, E.~Chu, N.~Parikh, and S.~Boyd, ``{SCS}: Splitting conic solver, version 3.2.3,'' \url{https://github.com/cvxgrp/scs}, Nov. 2022.

\bibitem{o2016conic}
B.~O’donoghue, E.~Chu, N.~Parikh, and S.~Boyd, ``Conic optimization via operator splitting and homogeneous self-dual embedding,'' \emph{Journal of Optimization Theory and Applications}, vol. 169, pp. 1042--1068, 2016.

\bibitem{shapiro2007tutorial}
A.~Shapiro and A.~Philpott, ``A tutorial on stochastic programming,'' \emph{Manuscript. Available at www2.isye.gatech.edu/ashapiro/publications.html}, vol.~17, 2007.

\bibitem{narayanan2020heterogeneity}
D.~Narayanan, K.~Santhanam, F.~Kazhamiaka, A.~Phanishayee, and M.~Zaharia, ``Heterogeneity-aware cluster scheduling policies for deep learning workloads,'' in \emph{Proceedings of the 14th USENIX Conference on Operating Systems Design and Implementation}, 2020, pp. 481--498.

\bibitem{serafini2014accordion}
M.~Serafini, E.~Mansour, A.~Aboulnaga, K.~Salem, T.~Rafiq, and U.~F. Minhas, ``Accordion: Elastic scalability for database systems supporting distributed transactions,'' \emph{Proceedings of the VLDB Endowment}, vol.~7, no.~12, pp. 1035--1046, 2014.

\bibitem{8675643}
H.~Nguyen and H.~La, ``Review of deep reinforcement learning for robot manipulation,'' in \emph{IEEE International Conference on Robotic Computing (IRC)}, 2019, pp. 590--595.

\bibitem{chen2022hierarchical}
J.~Chen, T.~Lan, and V.~Aggarwal, ``Option-aware adversarial inverse reinforcement learning for robotic control,'' in \emph{{IEEE} International Conference on Robotics and Automation}.\hskip 1em plus 0.5em minus 0.4em\relax {IEEE}, 2023.

\bibitem{chen2023multi}
J.~Chen, D.~Tamboli, T.~Lan, and V.~Aggarwal, ``Multi-task hierarchical adversarial inverse reinforcement learning,'' in \emph{Proceedings of the 40th International Conference on Machine Learning}, 2023.

\bibitem{hu2019interaction}
Y.~Hu, A.~Nakhaei, M.~Tomizuka, and K.~Fujimura, ``Interaction-aware decision making with adaptive strategies under merging scenarios,'' in \emph{2019 IEEE/RSJ International Conference on Intelligent Robots and Systems (IROS)}.\hskip 1em plus 0.5em minus 0.4em\relax IEEE, 2019, pp. 151--158.

\bibitem{al2019deeppool}
A.~O. Al-Abbasi, A.~Ghosh, and V.~Aggarwal, ``Deeppool: Distributed model-free algorithm for ride-sharing using deep reinforcement learning,'' \emph{IEEE Transactions on Intelligent Transportation Systems}, vol.~20, no.~12, pp. 4714--4727, 2019.

\bibitem{chen2021deepfreight}
J.~Chen, A.~K. Umrawal, T.~Lan, and V.~Aggarwal, ``Deepfreight: A model-free deep-reinforcement-learning-based algorithm for multi-transfer freight delivery,'' in \emph{Proceedings of the International Conference on Automated Planning and Scheduling}, vol.~31, 2021, pp. 510--518.

\bibitem{ye2015multi}
D.~Ye, M.~Zhang, and Y.~Yang, ``A multi-agent framework for packet routing in wireless sensor networks,'' \emph{sensors}, vol.~15, no.~5, pp. 10\,026--10\,047, 2015.

\bibitem{geng2023reinforcement}
N.~Geng, Q.~Bai, C.~Liu, T.~Lan, V.~Aggarwal, Y.~Yang, and M.~Xu, ``A reinforcement learning framework for vehicular network routing under peak and average constraints,'' \emph{IEEE Transactions on Vehicular Technology}, 2023.

\bibitem{gonzalez2023asap}
G.~Gonzalez, M.~Balakuntala, M.~Agarwal, T.~Low, B.~Knoth, A.~W. Kirkpatrick, J.~McKee, G.~Hager, V.~Aggarwal, Y.~Xue \emph{et~al.}, ``Asap: A semi-autonomous precise system for telesurgery during communication delays,'' \emph{IEEE Transactions on Medical Robotics and Bionics}, vol.~5, no.~1, pp. 66--78, 2023.

\bibitem{pac}
H.~Zhou, T.~Lan, and V.~Aggarwal, ``Pac: Assisted value factorization with counterfactual predictions in multi-agent reinforcement learning,'' in \emph{Advances in Neural Information Processing Systems}, vol.~35.\hskip 1em plus 0.5em minus 0.4em\relax Curran Associates, Inc., 2022, pp. 15\,757--15\,769.

\bibitem{gadiraju2023optimization}
D.~S. Gadiraju, V.~Lalitha, and V.~Aggarwal, ``An optimization framework based on deep reinforcement learning approaches for prism blockchain,'' \emph{IEEE Transactions on Services Computing}, 2023.

\bibitem{winstein2013tcp}
K.~Winstein and H.~Balakrishnan, ``Tcp ex machina: Computer-generated congestion control,'' \emph{ACM SIGCOMM Computer Communication Review}, vol.~43, no.~4, pp. 123--134, 2013.

\bibitem{mao2016resource}
H.~Mao, M.~Alizadeh, I.~Menache, and S.~Kandula, ``Resource management with deep reinforcement learning,'' in \emph{Proceedings of the 15th ACM workshop on hot topics in networks}, 2016, pp. 50--56.

\bibitem{luan2019sched2}
Y.~Luan, X.~Chen, H.~Zhao, Z.~Yang, and Y.~Dai, ``Sched$^2$: Scheduling deep learning training via deep reinforcement learning,'' in \emph{2019 IEEE Global Communications Conference (GLOBECOM)}.\hskip 1em plus 0.5em minus 0.4em\relax IEEE, 2019, pp. 1--7.

\bibitem{li2019deepjs}
F.~Li and B.~Hu, ``Deepjs: Job scheduling based on deep reinforcement learning in cloud data center,'' in \emph{Proceedings of the 4th International Conference on Big Data and Computing}, 2019, pp. 48--53.

\bibitem{zhang2020learning}
H.~Zhang, X.~Geng, and H.~Ma, ``Learning-driven interference-aware workload parallelization for streaming applications in heterogeneous cluster,'' \emph{IEEE Transactions on Parallel and Distributed Systems}, vol.~32, no.~1, pp. 1--15, 2020.

\bibitem{swarup2021task}
S.~Swarup, E.~M. Shakshuki, and A.~Yasar, ``Task scheduling in cloud using deep reinforcement learning,'' \emph{Procedia Computer Science}, vol. 184, pp. 42--51, 2021.

\end{thebibliography}


\end{document}